\documentstyle{mn}
\newif\ifAMStwofonts
\AMStwofontstrue
\input epsfig.sty
\input psfig.sty
\ifoldfss

  \ifCUPmtlplainloaded \else
    \NewTextAlphabet{textbfit} {cmbxti10} {}
    \NewTextAlphabet{textbfss} {cmssbx10} {}
    \NewMathAlphabet{mathbfit} {cmbxti10} {} 
    \NewMathAlphabet{mathbfss} {cmssbx10} {} 
  \fi
  \ifAMStwofonts
    \ifCUPmtlplainloaded \else
      \NewSymbolFont{upmath} {eurm10}
      \NewSymbolFont{AMSa} {msam10}
      \NewMathSymbol{\upi}     {0}{upmath}{19}
      \NewMathSymbol{\umu}     {0}{upmath}{16}
      \NewMathSymbol{\upartial}{0}{upmath}{40}
      \NewMathSymbol{\leqslant}{3}{AMSa}{36}
      \NewMathSymbol{\geqslant}{3}{AMSa}{3E}

      \let\geq=\geqslant 
    \fi
  \fi
\fi
\ifnfssone
  \newmathalphabet{\mathit}
  \addtoversion{normal}{\mathit}{cmr}{m}{it}
  \addtoversion{bold}{\mathit}{cmr}{bx}{it}

  \newmathalphabet{\mathbfit} 
  \addtoversion{normal}{\mathbfit}{cmr}{bx}{it}
  \addtoversion{bold}{\mathbfit}{cmr}{bx}{it}
  \newmathalphabet{\mathbfss} 
  \addtoversion{normal}{\mathbfss}{cmss}{bx}{n}
  \addtoversion{bold}{\mathbfss}{cmss}{bx}{n}
  \ifAMStwofonts
    \ifCUPmtlplainloaded \else
      \UseAMStwoboldmath
      \makeatletter
      \new@mathgroup\upmath@group
      \define@mathgroup\mv@normal\upmath@group{eur}{m}{n}
      \define@mathgroup\mv@bold\upmath@group{eur}{b}{n}
      \edef\UPM{\hexnumber\upmath@group}
      \new@mathgroup\amsa@group
      \define@mathgroup\mv@normal\amsa@group{msa}{m}{n}
      \define@mathgroup\mv@bold\amsa@group{msa}{m}{n}
      \edef\AMSa{\hexnumber\amsa@group}
      \makeatother
      \mathchardef\upi="0\UPM19
      \mathchardef\umu="0\UPM16
      \mathchardef\upartial="0\UPM40
      \mathchardef\leqslant="3\AMSa36
      \mathchardef\geqslant="3\AMSa3E

      \let\geq=\geqslant 
    \fi
  \fi
\fi
\ifnfsstwo

  \DeclareMathAlphabet{\mathbfit}{OT1}{cmr}{bx}{it}
  \SetMathAlphabet\mathbfit{bold}{OT1}{cmr}{bx}{it}
  \DeclareMathAlphabet{\mathbfss}{OT1}{cmss}{bx}{n}
  \SetMathAlphabet\mathbfss{bold}{OT1}{cmss}{bx}{n}
  \ifAMStwofonts
    \ifCUPmtlplainloaded \else
      \DeclareSymbolFont{UPM}{U}{eur}{m}{n}
      \SetSymbolFont{UPM}{bold}{U}{eur}{b}{n}
      \DeclareSymbolFont{AMSa}{U}{msa}{m}{n}
      \DeclareMathSymbol{\upi}{0}{UPM}{"19}
      \DeclareMathSymbol{\umu}{0}{UPM}{"16}
      \DeclareMathSymbol{\upartial}{0}{UPM}{"40}
      \DeclareMathSymbol{\leqslant}{3}{AMSa}{"36}
      \DeclareMathSymbol{\geqslant}{3}{AMSa}{"3E}

      \let\geq=\geqslant 
    \fi
  \fi
\fi
\ifCUPmtlplainloaded \else
  \ifAMStwofonts \else 
    \def\upi{\pi}
    \def\umu{\mu}
    \def\upartial{\partial}
  \fi
\fi

\def\simless{\mathbin{\lower 1pt\hbox
   {$\spose{\raise 5pt\hbox{$\char'074$}}\char'430$}}}
\def\simgreat{\mathbin{\lower 1pt\h   {$\spose{\raise 5pt\hbox{$\char'076$}}\char'430$}}}
\def\simgreat{\gapp}
\def\simless{\lapp}
\def\lapp{\mathbin{\raise2pt \hbox{$<$} \hskip-9pt \lower4pt \hbox{$\sim$}}}
\def\gapp{\mathbin{\raise2pt \hbox{$>$} \hskip-9pt \lower4pt \hbox{$\sim$}}}

\title{On the magnetic acceleration and collimation  \protect 
of astrophysical outflows}

\author[S. ~Bogovalov \& K. ~Tsinganos]
       {S. ~Bogovalov$^1$\thanks{Email: bogoval@photon.mephi.ru (SB); 
tsingan@physics.uch.gr (KT)} 
and K. ~Tsinganos$^{2,3{\Large \star}}$\\
 $^1$Astrophysics Institute of Moscow State Engineering
 Physics Institute, Moscow, 115409, Russia\\
 $^2$Department of Physics, University of Crete, GR-710 03 Heraklion, Crete,
 GREECE\\
 $^3$Foundation for Research and Technology Hellas (FORTH), GR-711 10 
 Heraklion, Crete, GREECE
 }

\date{Accepted 1998 November 29. Received September 11, 1998; 
in original form 1998 April 29}

\pagerange{\pageref{firstpage}--\pageref{lastpage}}
\pubyear{1994}

\begin{document}

\maketitle

\label{firstpage}

\begin{abstract}
The axisymmetric 3-D MHD outflow of cold plasma from a magnetized and 
rotating astrophysical object is numerically simulated with the purpose
of investigating the outflow's magnetocentrifugal acceleration and eventual
collimation.  Gravity and thermal pressure are neglected while a 
split-monopole is used to describe the initial magnetic field configuration. 
It is found that the stationary final state depends critically on a single 
parameter $\alpha$ expressing the ratio of the corotating speed at the Alfv\'en
distance to the initial flow speed along the initial monopole-like magnetic 
fieldlines.
Several angular velocity laws have been used for relativistic and
nonrelativistic outflows.
The acceleration of the flow is most effective at the equatorial plane and
the terminal flow speed depends linearly on $\alpha$.
Significant flow collimation is found in nonrelativistic
efficient magnetic rotators corresponding to relatively large values of
$\alpha \simgreat 1$ while very weak collimation occurs
in inefficient magnetic rotators with smaller values of $\alpha < 1$.
Part of the flow around the rotation and magnetic axis is cylindrically
collimated while the remaining part obtains radial asymptotics.
The transverse radius of the jet is inversely proportional to $\alpha$ while
the density in the jet grows linearly with $\alpha$.
For $\alpha \simgreat 5$ the magnitude of the flow in the jet remains
below the fast MHD wave speed everywhere.
In relativistic outflows, no collimation is found in the supersonic
region for parameters typical for radio pulsars. All above results 
verify the main conclusions of general theoretical studies on the magnetic 
acceleration and collimation of outflows from magnetic rotators and 
extend previous numerical simulations to large stellar distances.
\end{abstract}


\section{Introduction}

Plasma outflows from the environment of stellar or galactic objects,
in the form of collimated jets is a widespread phenomenon in astrophysics.
The most dramatic illustration of such highly collimated outflows may be
perhaps found in relatively nearby regions of star formation;
for example, in the Orion Nebula alone the Hubble Space Telescope (HST)
has observed hundreds of aligned Herbig-Haro objects (O'Dell \& Wen, 1994).
In particular, recent HST observations
show that several jets from young stars are highly
collimated within 30 - 50 AU from the source star with jet widths
of the order of tens of AU, although their initial opening angle
is rather large, e.g.,
$> 60^o$,
Ray et al (1996).
There is also a long catalogue of jets associated with AGN and possibly
supermassive black holes (Jones and Wehrle 1994, Biretta 1996).
To a less extend,
jets are also associated with older mass losing stars and planetary
nebulae, (Livio 1997), symbiotic stars (Kafatos 1996), black hole X-ray
transients (Mirabel \& Rodriguez 1996), supersoft X-ray sources (Kahabka \&
Trumper 1996), low- and high-mass X-ray binaries 
and cataclysmic variables (Shahbaz et al 1997).
Even for the two spectacular rings seen with the HST in SN87A, it has been
proposed that they may be inscribed by two precessing jets from an object
similar to SS433 on a hourglass-shaped cavity which has been created by
nonuniform winds of the progenitor star (Burderi and King, 1995,
Burrows et al 1995).

In the theoretical front, the morphologies of collimated outflows
have been studied, to a first approximation, in the framework of ideal
stationary or time-dependent magnetohydrodynamics (MHD). In {\it stationary}
studies, after the pioneering 1-D (spherically symmetric) works of Parker
(1963), Weber \& Davis (1967) and Michel (1969), it was Suess (1972) and
Nerney \& Suess (1975) who first modelled the 2-D (axisymmetric) interaction
of magnetic fields with rotation in stellar winds, by a
linearisation of the MHD equations in inverse Rossby numbers.
Although their perturbation expansion is not uniformly convergent but
diverges at infinity, they found a poleward deflection of the streamlines
of the solar wind caused by the toroidal magnetic field.
Blandford \& Payne (1982) subsequently demonstrated
that astrophysical jets may be accelerated magnetocentrifugally from
Keplerian accretion disks, {\it if} the poloidal fieldlines are inclined
by an angle of 60$^o$, or less, to the disk midplane (but see also,
Contopoulos \& Lovelace (1994), Shu et al, 1994, Cao, 1997, 
Meier et al 1997). This study introduced the
"bead on a rigid wire" picture, although these solutions are limited by the
fact that they contain singularities along the system's axis and also
terminate at finite heights above the disk.
Sakurai (1985) extended the Weber \& Davis
(1969) equatorial solution to all space around the star by iterating
numerically between the Bernoulli and transfield equations; thus, 
a polewards deflection of the poloidal fieldlines was found not 
only in an initially
radial magnetic field geometry, but also in a split-monopole one
appropriate to disk-winds, Sakurai (1987).
The methodology of meridionally self-similar exact MHD solutions with a
variable polytropic index was first introduced by
Low \& Tsinganos (1986) and Tsinganos \& Low (1989) in an effort to model
the {\it heated} axisymmetric solar wind.
Heyvaerts \& Norman (1989) have shown analytically that the asymptotics 
of a particular fieldline in non isothermal 
polytropic outflows is parabolic if it does not enclose a net current
to infinity; and, if a fieldline exists which does enclose a net
current to infinity, then, somewhere in the flow there exists a cylindrically
collimated core.
Later, Bogovalov (1995) showed analytically
that there {\it always} exists a fieldline in the outflowing part of a
{\it rotating} magnetosphere which encloses a finite total poloidal current
and therefore the asymptotics of the outflow {\it always} contains a
cylindrically collimated core.
In that connection, it has been shown in Bogovalov (1992) that the
poloidal fieldlines are
deflected towards the polar axis for the split monopole geometry and
relativistic or nonrelativistic speeds of the outflowing plasma.
Sauty \& Tsinganos (1994) have self-consistently determined the shape
of the fieldlines from the base of the outflow to infinity for
nonpolytropic cases and provided a simple
{\it criterion} for the transition of their asymptotical shape from conical
(in inefficient magnetic rotators) to cylindrical (in efficient magnetic
rotators). They have also conjectured that as a young star spins down loosing
angular momentum, its collimated jet-type outflow becomes gradually a
conically expanding wind. Nevertheless, the degree of the collimation of
the solar wind at {\it large} heliocentric distances
remains still observationally unconfirmed, since
spacecraft observations still offer ambiguous evidence on this question.
Another interesting property of collimated outflows has emerged from
studies of various self-similar solutions, namely that in a large portion of
them cylindrical collimation is obtained only after some oscillations
of decaying amplitude in the jet-width appear (Vlahakis \& Tsinganos 1997).
Radially self-similar models with cylindrical asymptotics
for self-collimated and magnetically dominated outflows from accretion
disks have been constructed in Ostriker (1997).
All existing cases of self-similar, jet- or, wind-type exact
MHD solutions can be unified by a systematic analytical treatment wherein
the available today examples of exact solutions emerge as special cases of a
general formulation while at the same time new families with various
asymptotical shapes, with (or without) oscillatory behaviour 
emerge as a byproduct of this systematic method (Vlahakis \& Tsinganos, 1998).
Altogether, some general trends on the behaviour of stationary, analytic,
axisymmetric MHD solutions for MHD outflows seem to be well at hand.

However, observations seem to indicate that jets may inherently be variable.
Thus, time-dependent simulations may be useful for a detailed
comparison with the observations.
Uchida \& Shibata (1985) were the
first to perform time-dependent simulations and demonstrate that a
vertical disk magnetic field if twisted by the rotation of the disk
can launch bipolar plasma ejections through the torsional
Alfv\'en waves it generates. However, this mechanism applies to fully
episodic plasma ejections and no final stationary state is reached to be
compared with stationary studies.
Similar numerical
simulations of episodic outflows from Keplerian disks driven by torsional
Alfv\'en waves on an initially vertical magnetic field have been presented
by Ouyed \& Pudritz (1997a,b).
Goodson et al (1997) have proposed a time-dependent jet launching and
collimating mechanism which produces a two-component outflow: hot, well
collimated jet around the rotation axis and a cool but slower disk-wind.
Stationary MHD jet-type outflows have been found in the study of Romanova
et al (1997), as the asymptotic state of numerical simulations wherein a
small initial velocity is given in the plasma in a tapered monopole-like
magnetic field. Numerical viscosity however results in nonparallel flow
and magnetic fields in the poloidal plane in the limited grid space of
integration.
Washimi \& Shibata (1993) modelled axisymmetric
thermo-centrifugal winds
with a {\it dipole} magnetic flux distribution $B_p^2(\theta) \propto
(3 \cos^2 \theta +1 )$ on the stellar surface (and a radial field in
Washimi 1990). In
this case the magnetic pressure distribution varies approximately as
$B^2_\phi \propto  B_p ^2 \sin^2 \theta $
such that it has a maximum at about $cos^{-1} 2\theta_o \approx -1/3$, or,
$\theta_0 \approx 55^o$. As a result, the flow and flux is directed
towards the pole and the equator from the midlatitudes around $\theta_o$.
The study was performed for uniform in latitude rotation rates and up to
60 solar radii in the equatorial plane.
Bogovalov (1996, 1997)
modelled numerically the effects of the Lorentz force
in accelerating  and collimating a cold plasma with an initially
monopole-type magnetic field, in a region limited also by
computer time, i.e., the near zone to the central spherical object.

This paper presents an extension of the previous results to {\it large} 
distances from the star by using a new method for the continuation of
the small simulation box solution to very large distances.
It also examines the efficiency of magnetic rotators of various strengths
in transforming rotational energy to directed kinetic energy and how a wide
range of rotation rates affects the poloidal geometry of a magnetic field.
In order to achieve these objectives, the rather complicated nature of the 
problem requires that we start by limiting the investigation 
to the simplest model of cold plasma flow in a uniform in latitude
monopole magnetic field along which there is an outflow with velocity
$V_o$, as the initial condition. This simplification allows us 
to better study and understand the nonlinear effects of the
magnetocentrifugal forces alone in shaping the final stationary
configuration. 
At the same time, however, it does not allow us to perform a direct comparison 
of the obtained results with observations. Clearly this is not the goal  
of the present paper since for such a comparison we need to include 
gravity and thermal pressure in our computations. 
Such a study has been already performed in the context of the solar 
wind and it will be presented elsewhere.

The paper is organised as follows. In Secs. 2 and 3, the initial configuration
used together with the method for the numerical simulation in the nearest
zone is discussed. In Secs. 4 and 5 the analytical method for extending the 
integration to unlimited large distances outside the near zone are briefly 
described.  
In Secs. 6 and 7 we discuss the results in the near zone containing the 
critical surfaces and in the asymptotic regime of the collimated outflow, 
for a uniform rotation.
In Sec. 8 a rotation law appropriate for an accretion disk is
used, while in Sec. 9 we briefly discuss results of a relativistic
modelling. A brief summary is finally given in the last Section 9.

\section{The model of a rotator with a monopole-like magnetic field and 
the objectives of this paper}

A general analysis of the asymptotical properties of nonrelativistic or 
relativistic magnetized winds has been already performed, e.g., Heyvaerts 
and Norman (1989), Chiueh et al (1991), Bogovalov (1995). The main conclusions 
from such studies can be summarized as follows: 

\begin{enumerate}
\item At large distances from the central source the poloidal magnetic field 
is similar (although not exactly the same) to a split-monopole field, 

\item there exist cylindrically collimated and radially expanding field 
lines, while the total electric current enclosed by any magnetic surface 
is nonzero, 

\item several physical quantities accross the jet can be expressed 
by simple formulas, under certain conditions.\\
\end{enumerate}
The model of an axisymmetric rotator with a monopole-like magnetic field
was first used by Michel (1969) for the investigation of the cold
plasma flow in the absense of gravity in a prescribed poloidal magnetic field.
Later this model was used by  Sakurai (1985) in an attempt to solve
selfconsistently the problem of the nonrelativistic plasma outflow from a 
stellar object. 
This model may be used to the study of plasma outflow from the
magnetospheres of various cosmic objects under the following conditions:\\
\begin{enumerate}
\item[(1)]  We are interested in the plasma flow at {\it large} 
distances where all
magnetic field lines are open. In this case, the field of any axisymmetric 
rotator, no matter what is the nature of the central object, becomes the 
field of the so-called split-monopole in the far zone. 
In other words, the condition that should be fulfilled is that the distance 
to the central source should be much larger than all the dimensions of the 
central source (see also Heyvaerts \& Norman 1989, Bogovalov 1995).
The solution for the split monopole field can be easily obtained from
a monopole-like solution by a simple reversing of the magnetic field in 
one of the hemispheres. \\

\item[(2)]  In considering the plasma flow at such large distances it 
is natural to neglect gravity. Thermal pressure however may play an important 
role even at large distances from the central object (Bogovalov 1995). But
in this paper we are interested to isolate the effects arising purely
from the magnetic field. Thus, to make our analysis as simple as possible,  
in this paper we neglect thermal pressure too.\\

\item[(3)] We are also interested to study the magnetocentrifugal   
acceleration of the plasma. 
This acceleration process can be studied with a monopole like magnetic field 
regarded as a first approximation to the more realistic acceleration in a 
dipole-like magnetic field with open field lines.  
Gravity and thermal pressure can be neglected in this case (Michel 1969).\\
\end{enumerate}
The previous claims (i) - (iii) are the result of a rather general analysis. 
Among the main goals of the present work is to verify these general conclusions 
in the context of the split monopole model and assumptions (1) - (3).

\section{The problem in the nearest zone.}

To obtain a stationary solution of the problem in the nearest zone of the
star containing the 
critical surfaces, it is
needed to solve the complete system of the time-dependent MHD equations
and look for an asymptotic stationary state. In order to isolate the effects
of the magnetic field in determining the shape of the streamlines, we
shall neglect gravity and thermal pressure gradients, as discussed above. 
With these
simplifications, the flow of the nonrelativistic plasma is described by the
set of the familiar MHD equations,
\smallskip
\begin{equation}
{\bf B_{p}}={\nabla\psi\times {\bf \hat {\varphi}}\over r}.
\label{psi}
\,,
\end{equation}
\smallskip
\begin{equation}
{\partial\psi\over\partial t} =-V_{r}{\partial\psi\over\partial
r}-V_{z}{\partial\psi\over\partial z}
\,,
\end{equation}
\smallskip
\begin{equation}
{\partial \rho \over\partial t}=- {1\over r} {\partial\over \partial
r} (\rho rV_{r}) - {\partial \over \partial z}( \rho V_{z})
\,,
\end{equation}
\smallskip
\begin{equation}
{\partial B_{\varphi}\over\partial t}={\partial
\over\partial z} (V_{\varphi}B_{z}-V_{z}B_{\varphi})
-{\partial \over\partial r} (V_{r}B_{\varphi}-V_{\varphi}B_{r})
\,,
\end{equation}
\smallskip
\begin{eqnarray}
{\partial V_{\varphi}\over\partial
t} & = &-{V_{r}\over r} {\partial \over \partial r}(rV_{\varphi})
-V_{z}{\partial V_{\varphi}\over\partial z}+\nonumber \\
& & {1\over 4 \pi \rho}\left( B_{r}{\partial \over r\partial r}(rB_{\varphi})
+B_{z} {\partial B_{\varphi}\over\partial z}\right) 
\,,
\end{eqnarray}
\smallskip
\begin{eqnarray}
{\partial V_{z}\over\partial t} & = & -V_{r}{\partial V_{z}\over\partial
r}-V_{z}{\partial V_{z}\over\partial z}-{1\over 8\pi \rho r^{2}}{\partial
\over\partial z} (rB_{\varphi})^{2} - \nonumber \\
& & {B_{r}\over 4\pi \rho}\left({\partial B_{r}\over\partial z}-{\partial
B_{z}\over\partial r}\right) 
\,,
\end{eqnarray}
\smallskip
\begin{eqnarray}
{\partial V_{r}\over\partial t} & = & -V_{r}{\partial V_{r}\over\partial
r}-V_{z}{\partial V_{r}\over\partial z}-{1\over 8\pi \rho r^{2}}{\partial
\over\partial r} (rB_{\varphi})^{2} + \nonumber \\
& & {V_{\varphi}^{2}\over r}+
{B_{z}\over 4\pi \rho}\left({\partial B_{r}\over\partial z}-{\partial
 B_{z}\over\partial r}\right) 
\,,
\end{eqnarray}

\noindent
where we have used cylindrical coordinates $(z, r, \varphi )$,
$\rho$ is the density, $\vec V$ the flow field and $\vec B$ the magnetic
field with a poloidal magnetic flux denoted by $\psi (z, r)$.

A correct solution of the problem requires a specification of the
appropriate boundary conditions at some spherical boundary $R=R_o$
of the integration.  \\
1. A constant plasma density $\rho_o$ at $R=R_o$.\\
2. A constant total plasma speed $V_o$ in the corotating frame of
reference at $R=R_o$,
$V_{(r,o)}^2 + V_{(z,o)}^2 + (V_{(\varphi , o)} - \Omega r_o)^2 = V_o^2$\\
3. A constant and uniform in latitude distribution of the magnetic flux
function $\psi = \psi_o$ at $R=R_o$.\\
4. Finally, the continuity of the tangential component of the electric field
across the stellar surface in the corotating frame gives the last condition,
$(V_{(\varphi , o)} - \Omega r_o) B_{(p, o)} - V_{(p, o)}B_{(\varphi , o)}
= 0$.

We shall use dimensionless variables,  $Z=z/R_{a}$, $X=r/R_{a}$,
$\tau=tV_{0}/R_{a}$ where $R_a$ is the Alfv\'en spherical radius of an 
initially radial, monopole-like, nonrotating  magnetic field and define the
dimensionless parameter 
\begin{equation}
\alpha={\Omega R_{a}\over V_{0}}
\,.
\end{equation}
This parameter $\alpha$ characterizes the influence of the magnetic field and
rotation on the acceleration and collimation of the plasma. It is
proportional to the time the plasma spends in the subAlfvenic region in
each period of rotation.
Note that although the governing equations (1-7) do not depend on $\alpha$,
the final solution does depend on $\alpha$ through the boundary conditions
at the base of the integration $R=R_o$ (condition 4).

As we shall see, the solution in the nearest zone will relax after sufficient
time to a stationary state. This stationary state will be next used as the
input for specifying the boundary conditions in the super-fast magnetosonic
region. In this way we shall be able to obtain a complete solution of the
stationary problem, from the base up to large distances downstream.

\section{The stationary problem}

Below we shall consider that the plasma may be relativistic, or
nonrelativistic and as before we shall neglect gravity and thermal pressure.
By $U_{p}=\gamma V_p/c$ and $U_{\varphi}=\gamma V_{\varphi}/c$, we denote the
poloidal and azimuthal 4-speeds where $V_{\varphi}$ is the
azimuthal and $V_p$ the poloidal components of the velocity while
$\gamma$ is the Lorentz factor of the plasma.
In the following subsection we review the basic quantities which remain
invariant along a poloidal streamline $\psi = const.$ and express momentum
balance along such a poloidal streamline. In the next subsection we adopt a
new coordinate system for dealing with the transfield equation expressing
momentum balance across the poloidal streamlines.

\subsection{MHD Integrals}

As is well known, the stationary MHD equations admit four integrals. They
are:
\begin{description}
\item[($\alpha$)] The ratio of the poloidal magnetic and
mass fluxes,  $c F(\psi )$
\begin{equation}
cF (\psi ) = {B_{p}\over 4\pi \rho V_{p}}
\,.
\end{equation}
\item[($\beta$)]  The total angular momentum per unit mass $L (\psi )$,
\begin{equation}
rcU_{\varphi}-cFrB_{\varphi}=L(\psi)
\,.
\end{equation}
\item[($\gamma$)] The corotation frequency $\Omega (\psi )$ in the
frozen-in condition
\begin{equation}
cU_{\varphi}B_{p} - cU_{p}B_{\varphi} = r \gamma B_{p} \Omega (\psi )
\,.
\label{freez}
\end{equation}
\noindent
\item[($\delta$)] The total energy $c^2W(\psi )$ in the equation for
total energy conservation,
\begin{equation}
\gamma c^2 - cF(\psi)r\Omega (\psi)B_{\varphi} = c^2 W (\psi)
\,.
\end{equation}
\end{description}

\subsection{The transfield equation in the coordinates $(\psi, \eta )$.}

Momentum balance across the poloidal fieldlines is expressed by the
transfield equation which determines and their shape. 
This is a rather complicated nonlinear partial differential
equation of mixed elliptic/hyperbolic type.
For analysing the behaviour of the plasma at large distances, it occured
to us that it is rather convenient to work with this transfield equation
in an orthogonal curvilinear coordinate system ($\psi, \eta$) formed by
the tangent to the poloidal magnetic field line $\hat \eta = \hat p$ and the
first normal towards the center of curvature of the poloidal lines,
$\hat \psi = \nabla \psi /|\nabla \psi|$ (Sakurai, 1990).
A geometrical interval in these coordinates can be expressed as
\begin{equation}
(d{\bf r})^{2}=g^2_{\psi}d\psi^{2}+g^2_{\eta}d\eta^{2}+r^{2}d\varphi^{2}
\,,
\end{equation}
where $g_{\psi}, g_{\eta}$ are the corresponding line elements or components
of the metric tensor.

If $T^{ij}$ is the energy-momentum tensor of the plasma flow ($\rho \vec V$) 
and electromagnetic field ($\vec E, \vec B$) (Landau \& Lifshitz 1975), 
the equation $\partial T^{\psi k}/\partial x^{k}=0$
(with covariant derivatives)
has the following form in the coordinates ($\psi, \eta$),
\begin{displaymath}
{\partial\over\partial\psi}\left[
{B^{2}-E^{2} \over 8\pi} \right] - {1\over r}{\partial r\over
\partial\psi}\left[\rho V_{\varphi}^{2} -
{B_{\varphi}^{2}-E^{2} \over 4\pi} \right]- 
\end{displaymath}
\begin{equation} 
{1\over g_{\eta}} {\partial g_{\eta} \over \partial\psi}
\left[\rho V_{p}^{2} -{B_{p}^{2}-E^{2} \over 4\pi}\right] = 0 \,.  
\label{transfield}
\end{equation}

The first term in this equation is the gradient of the pressure of the
electromagnetic field, while the second is the sum
of the inertial terms due to the motion of the plasma in the azimuthal
direction and also due to the tension of the toroidal magnetic field.
To better understand the physical meaning of the last term, note that
since $\hat \eta$ is perpendicular to $\hat \psi$ we have,

\begin{equation}
{1\over g_{\eta}} {\partial g_{\eta} \over \partial\psi} = {1\over
rR_{c}B_{p}},
\end{equation}

\noindent
where $R_{c}$  is the radius of curvature of the poloidal magnetic field
lines, with $R_{c}$ positive if the center of curvature is in the domain
between the line and the axis of rotation and negative in the opposite case.

With this expression, Eq. (\ref{transfield}) becomes,
\begin{displaymath}
{\partial\over\partial\psi}\left[{B_{p}^{2}\over
8\pi}\right] +{1\over 8\pi r^{2}}{\partial\over
\partial \psi}\left[r^{2}(B^{2}_{\varphi}-E^{2})\right] - 
\end{displaymath}
\smallskip
\begin{equation}
{1\over 4\pi r}
{\partial r\over\partial\psi}
{U_{\varphi}^{2}B_{p}\over U_{p}F(\psi)}  
-{[U_{p}-F(\psi)(1-(r\Omega/c)^{2})B_{p}]\over 4\pi rR_{c}F(\psi)}=0 
\,. 
\label{trans2}
\end{equation}
\smallskip\noindent
From this equation, it may be seen that the last term is the sum of
the inertia of plasma and fields, connected with the motion of
the plasma and Poynting flux along the poloidal field line and
tension of the poloidal field line.

\subsection{The transfield equation for nonrelativistic plasmas with gravity
and thermal pressure included}

For completeness of the picture we present briefly here the transfield 
equation for a nonrelativistic plasma flow to demonstrate that our method of 
solution of the stationary problem in the hyperbolic region can be applied
directly to this case too.
The energy momentum conservation equation in the presence of gravity in the
nonrelativistic limit is modified as $\partial T^{\psi k}/\partial x^{k}= 
-\rho \partial \Phi/\partial x^{\psi}$, 
where $T^{\psi k}$ is again the energy-momentum tensor, $\Phi=-{GM/R}$
is the gravitational
potential of the star with mass $\it M$ and $G$ is the gravitational constant.
This equation has the following form in our curvilinear coordinates if some 
thermal pressure $P$ is also included,

\begin{displaymath}
{\partial\over\partial\psi}
\left[P+{B^2\over 8\pi}\right]
- {1\over r}{\partial r\over \partial\psi}
\left[\rho V_{\varphi}^{2}-{B_{\varphi}^{2}\over 4\pi}
\right]- 
\end{displaymath}
\smallskip
\begin{equation}
{1\over g_{\eta}} {\partial g_{\eta} \over \partial\psi}
\left[\rho V_{p}^{2}-{B_p^2\over 4\pi}\right] =
-\rho {\partial \Phi\over\partial\psi} \,.
\label{transfieldg}
\end{equation}

All other equations, describing the flow of plasma along
field lines will be the same except the energy conservation
equation which is modified as follows
\begin{equation}
{V^{2}\over 2}+{\delta \over \delta -1}{P\over \rho} + \Phi
-cF(\psi)r\Omega (\psi)B_{\varphi}=W(\psi)c^2
\,.
\end{equation}
where $\delta$ is the polytropic index of the plasma.

\section{The solution in the far zone.}

It is convenient to solve the transfield equation in the system of
the curvilinear coordinates introduced above. The unknown variables
are $z(\eta, \psi)$ and $r(\eta, \psi)$. Therefore we need to know the
quantities $g_{\eta}$, $g_{\psi}$,  $r_{\psi}$, $z_{\psi}$,
$r_{\eta}$,  $z_{\eta}$, where $r_{\eta}=\partial r/\partial\eta$, $z_{\eta}=
\partial z/\partial\eta$, $r_{\psi}=\partial r/\partial\psi$,
$z_{\psi}=\partial z/\partial\psi$.

First, the metric coefficient $g_{\eta}$ is obtained from the
transfield equation (\ref{transfield}),

\begin{equation}
 g_{\eta} =\exp{(\int\limits_0^\psi G(\eta,\psi)d\psi)}\,,
 \label{ga}
 \end{equation}
 where
\begin{displaymath}
G(\eta,\psi)  = 
\end{displaymath}
\smallskip
\begin{equation} 
\displaystyle{
{{\partial\over\partial\psi}\left[{1\over
        8\pi}(B^{2}-E^{2})\right] - {1\over r} {\partial r\over
        \partial\psi} \left[\rho V_{\varphi}^2-{1\over 4\pi}
        (B_{\varphi}^{2}-E^{2})\right] \over \left[ \rho V_{p}^2-{1\over
        4\pi}(B_{p}^{2}-E^{2})\right]}
}
\,,
\label{G}
\end{equation}
for a cold plasma.
For a nonrelativistic flow with finite thermal pressure and gravity
the function $G$ will have the form
\begin{displaymath}  
G(\eta,\psi)  = 
\end{displaymath}
\smallskip
\begin{equation}
\displaystyle{  
{{\partial\over\partial\psi}\left[{1\over
        8\pi}B^{2} +P\right] +{\partial\over\partial\psi}\Phi
        -{1\over r} {\partial r\over
        \partial\psi} \left[\rho V_{\varphi}^2-{1\over 4\pi}
        B_{\varphi}^{2}\right] \over \left[ \rho V^2_{p} - {1\over
        4\pi}B_{p}^{2}\right]}
}
\,.
\label{G2}
\end{equation}
The lower limit of the integration in Eq. (\ref{ga}) is chosen to be 0
such that the coordinate $\eta$ is uniquely defined.
In this way $\eta$ coincides with the coordinate $z$ where the
surface of constant $\eta$ crosses the axis of rotation.

Second, the metric coefficient $g_{\psi}$ is given in terms of the
magnitude of the poloidal magnetic field by,
\begin{equation}
g_{\psi} ={1\over rB_{p} }
\,.
\end{equation}

To obtain the expressions of $r_{\psi}$, $z_{\psi}$,
$r_{\eta}$,  $z_{\eta}$ we may use the orthogonality condition
\begin{equation}
 r_{\eta}r_{\psi}+z_{\eta}z_{\psi}=0
\,,
\label{orto}
\end{equation}
and also the fact that they
are related to the metric coefficients $g_{\eta}$ and $g_{\psi}$
as follows,
\begin{equation}
g^2_{\eta}=r_{\eta}^{2}+z_{\eta}^{2}
\,,
\label{gpsi}
\end{equation}
\begin{equation}
g^2_{\psi}=r_{\psi}^{2}+z_{\psi}^{2}
\,.
\label{galpha}
\end{equation}

Thus, by combining the condition of orthogonality (\ref{orto}) and equations
(\ref{gpsi}) and (\ref{galpha}) the remaining values of
$r_{\eta}$, $z_{\eta}$ are obtained,

\begin{equation}
r_{\eta}=-{z_{\psi} g_{\eta} \over g_{\psi}}
\,,
\label{xa}
\end{equation}

\begin{equation}
z_{\eta}={r_{\psi} g_{\eta} \over g_{\psi}}
\,.
\label{za}
\end{equation}
with $g_{\eta}$  calculated by the expression (\ref{ga}).
For the numerical solution of the system of equations (\ref{xa} - \ref{za})
a two step Lax-Wendroff method is used on a lattice with a dimension equal
to 1000.

Equations  (\ref{xa} - \ref{za}) should be supplemented by
appropriate boundary conditions on some initial surface of constant $\eta$. 
The equations for $r_{\psi}$, $z_{\psi}$
defining this initial surface in cylindrical coordinates are
as follows

\begin{equation}
{\partial r\over \partial\psi}={B_{z}\over rB_{p}^{2}}
\label{rpsi}
\,,
\end{equation}

\begin{equation}
{\partial z\over \partial\psi}=-{B_{r}\over rB_{p}^{2}}
\,.
\end{equation}
We need to specify  on this surface the integrals $F(\psi), L(\psi),
\Omega(\psi)$ and $W(\psi)$ as the boundary conditions for the initial 
value problem.
To specify the initial surface of constant $\eta$ and the above integrals, 
we use the results of the solution of
the problem in the nearest zone when a stationary solution is
obtained for the time-dependent problem.

\section{Results in the nearest zone for uniform rotation,
$\Omega ({\psi}) =\Omega_o$}

As the star starts rotating, in the initially radial magnetosphere
an MHD wave propagates outwards carrying the effect of the rotation and
deflecting the fieldlines polewards by the Lorentz force (Fig. 1).   

\begin{figure}
\centerline{\psfig{file=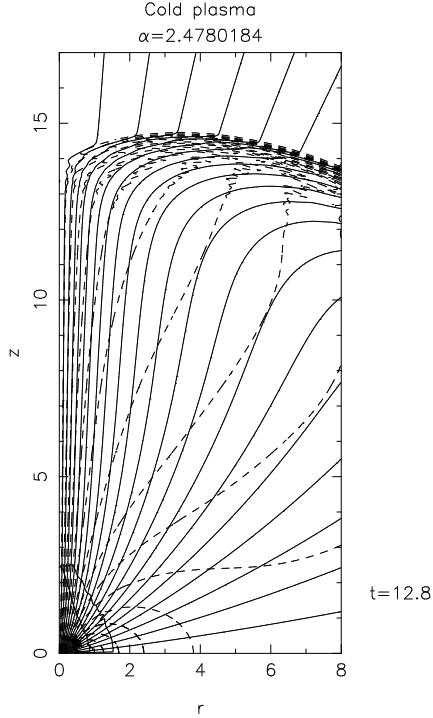,height=10.0truecm,angle=270}}
\caption{\it A near zone snapshot on the poloidal plane (Z, r)
from the simulation showing the
change of the shape of the poloidal magnetic fieldlines
from an initially uniform with latitude radial monopole and before
a stationary state is reached. Distances are given in units of the
Alfv\'en radius $R_a$ and time in units of the Alfv\'en crossing time
$t_a = R_a/V_a$}
\end{figure}

At times
sufficiently long after the shock wave has reached the end boundary
of the simulation ($t >> 1$), a final equilibrium state is reached
(Figs. 3).
In this stationary state, the poloidal magnetic field and the plasma
density are increased along the axis because of the focusing of the
field lines towards the pole.

In Figs. (3) the distances are given in units of the initial Alfvenic radius 
$R_a$.
Thick lines indicate Alfv\'en and fast critical surfaces while thin lines 
the poloidal magnetic field.  At $t=0$ the Alfv\'en surface
is spherical at $R=1$. Dotted lines indicate poloidal currents.

With the velocity maintaining the constant initial value $V_o$
along the Z-axis, the ratio $B_p/\rho$ remains constant along this axis
($\psi=0$). As a result, the Alfv\'en speed at a given point of the Z-axis,
$V_A=B_p/\sqrt{4\pi\rho}$, increases as the square root of the density
increases there because of focusing. It follows that the Alfv\'en
transition at $R_A(\theta=0)$ occurs further downstream where
$V_o(R_A) = V_A(R_A)$, i.e., at $R_A (\theta =0)> 1$.
As we move meridionally off the polar axis toward the equator on the
other hand, the bulk flow speed is increased because of the
magnetocentrifugal acceleration; it thus hits the Alfv\'en value
earlier than it does on the axis, i.e., $V(\theta) = V_A(\theta)$,
at $R_A (\theta)< R_A(\theta =0)$. Finally, on the equator
the flow speed increases rapidly with the result that the flow becomes
much earlier super Alfv\'enic there in comparison to the polar axis. The
degree of elongation of the Alfv\'en surfaces along the symmetry axis
increases with the value of $\alpha$.

The behaviour of the fast surface has similarities and differences with the
shape of the Alfv\'en surface. First, both these critical surfaces coincide
at $\theta = 0$.
As we move meridionally off the polar axis toward the equator on the
other hand, their shape becomes different. This is due to the contribution
of the azimuthal field, since in this case of cold plasma
$V_f^2 = (B_p^2 + B_\phi^2)/4\pi\rho $. Thus, for small colatitudes
$\theta$, $V_f$ increases due to the contribution of $B_\phi$ while also
$V$ slightly increases because of the magnetocentrifugal acceleration.
However, initially $V_f$ increases faster and therefore the
fast transition is postponed further downstream. At larger $\theta$
however, the competing flow speed $V$ increases faster than $V_f$ does
and the fast transition occurs closer and closer to the origin.

\setcounter{figure}{1}
\begin{figure}
\centerline{\psfig{file=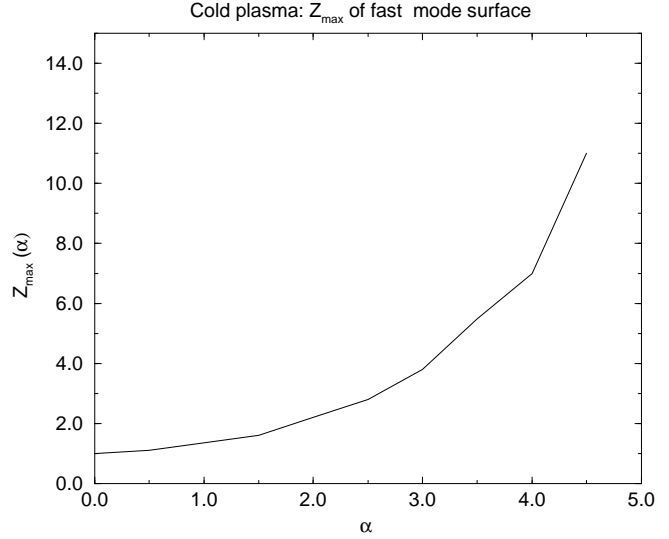,height=8.0truecm,angle=360}}
\caption{\it Maximum height $Z_{max}$ of fast critical surface
as a function of $\alpha$, in units of the base radius $R_a$.
}
\end{figure}

The dependance of the maximum height $Z_{max}$ of the fast critical surface
as a function of $\alpha$ is interesting too (Fig. 2). As $\alpha$ increases,
$Z_{max}$ increases rapidly. That means that for sufficiently fast rotators
with $\alpha \geq 5$, $Z_{max}$ goes to infinity and the flow may 
stay sub-fast all the way to large distances.  This result may have 
some important implications in the general theory of MHD winds.
It clearly indicates that a super-fast stationary solution may not be 
obtained for all sets of parameters. For large $\alpha$
no super-fast stationary state is found and thus the equilibrium is 
vulnerable to instabilities.  Such a situation may take place, 
for example, in young rapidly rotating stars. 
Outflows from classical accretion disks (Shakura \& Sunyaev 1973)  are 
also expected to have such large values of $\alpha$, since the Alfv\'en and 
sound
speeds in such disks are much smaller than their respective Keplerian speed.
Thus, since $V_a \sim V_s << V_K$ and $r_a > r_K$, we have
$$
\alpha = {\Omega r_a \over V_a} >> {\Omega r_K \over V_K} \approx 1
\,.
$$

\setbox31=\vbox{\hsize=10 truecm \vsize=10truecm
\psfig{figure=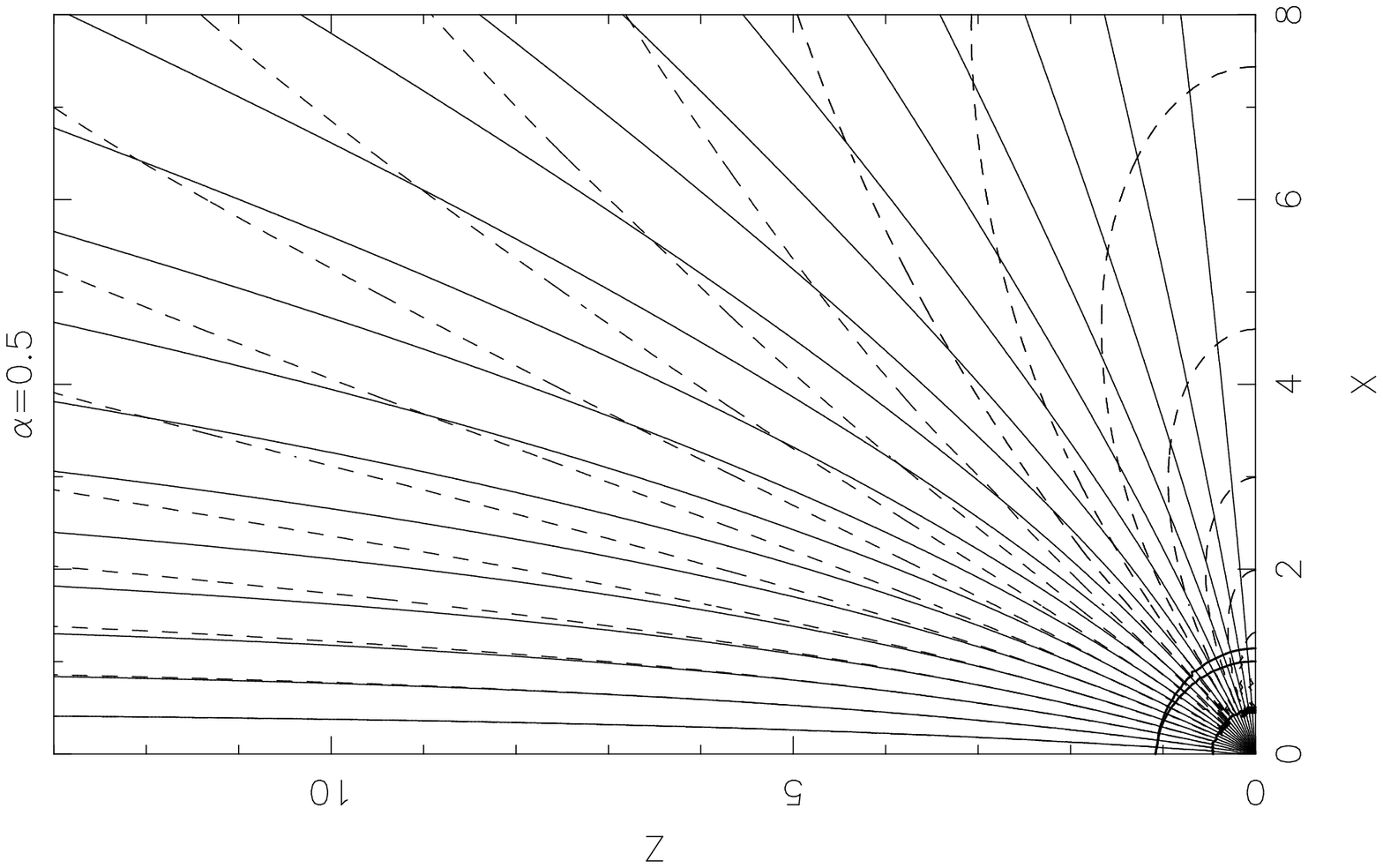,height=10.0truecm,angle=270}}
\setbox32=\vbox{\hsize=10 truecm \vsize=10truecm
\psfig{figure=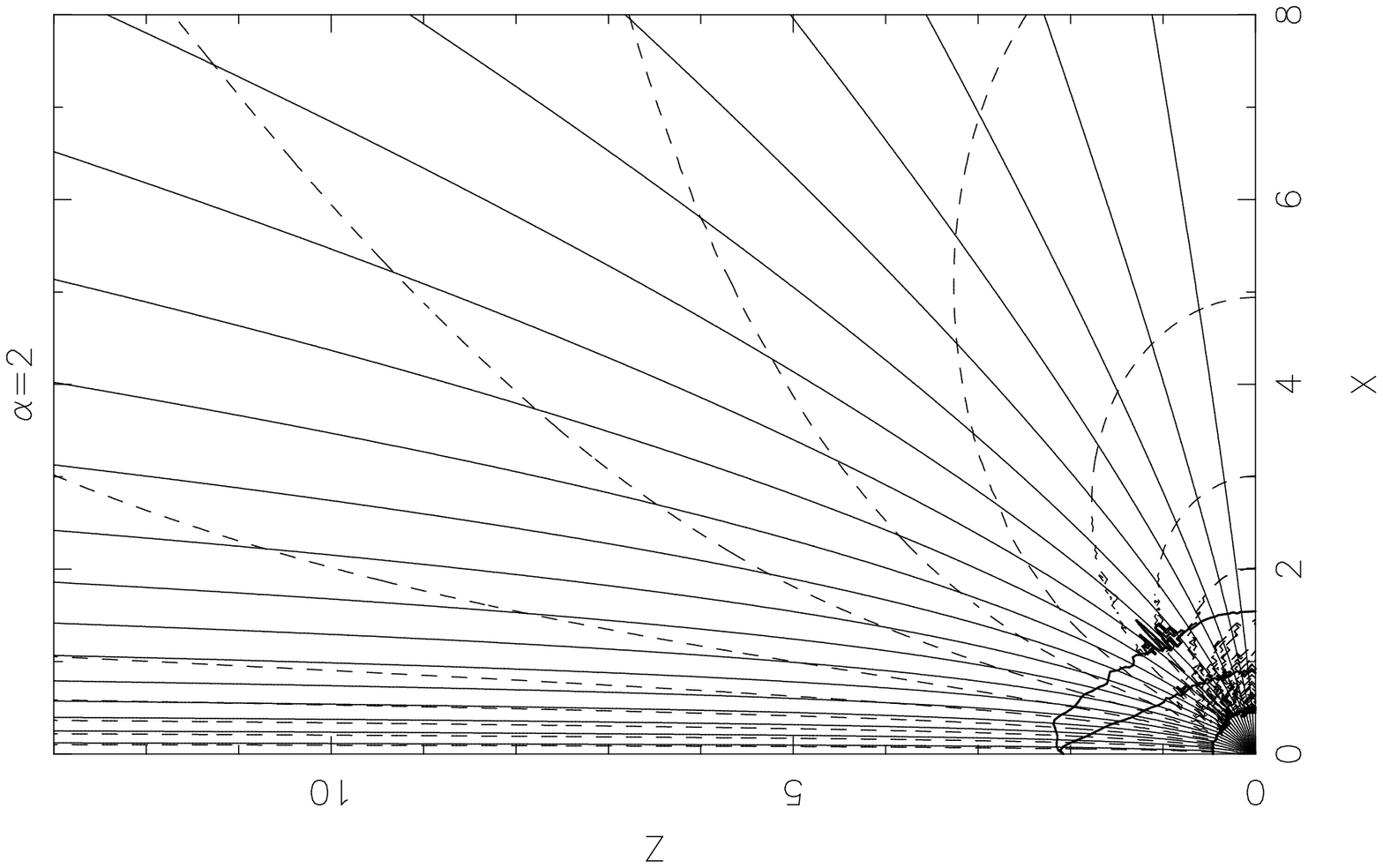,height=10.0truecm,angle=270}}
\setbox33=\vbox{\hsize=10 truecm \vsize=10truecm
\psfig{figure=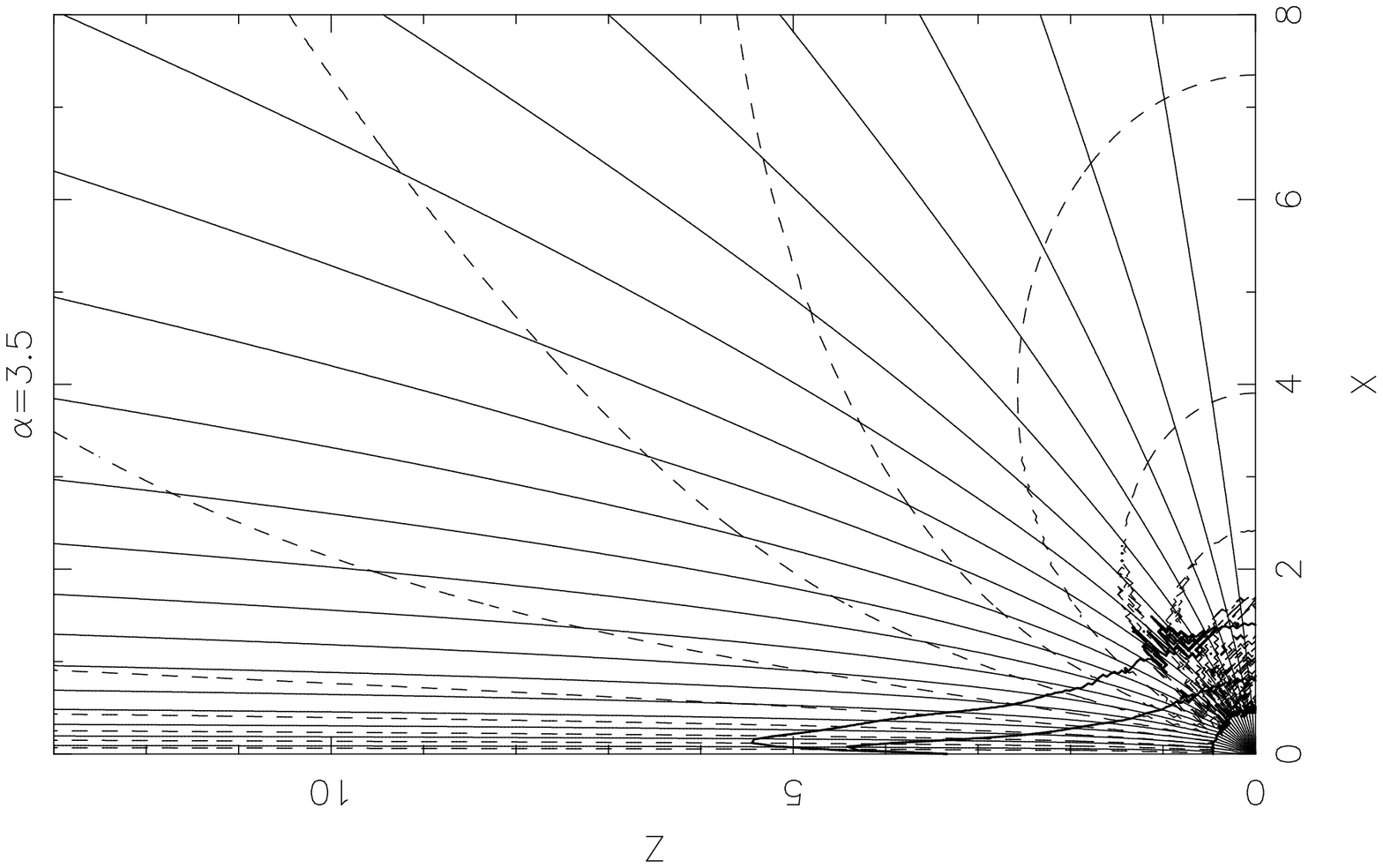,height=10.0truecm,angle=270}}
\setbox34=\vbox{\hsize=10 truecm \vsize=10truecm
\psfig{figure=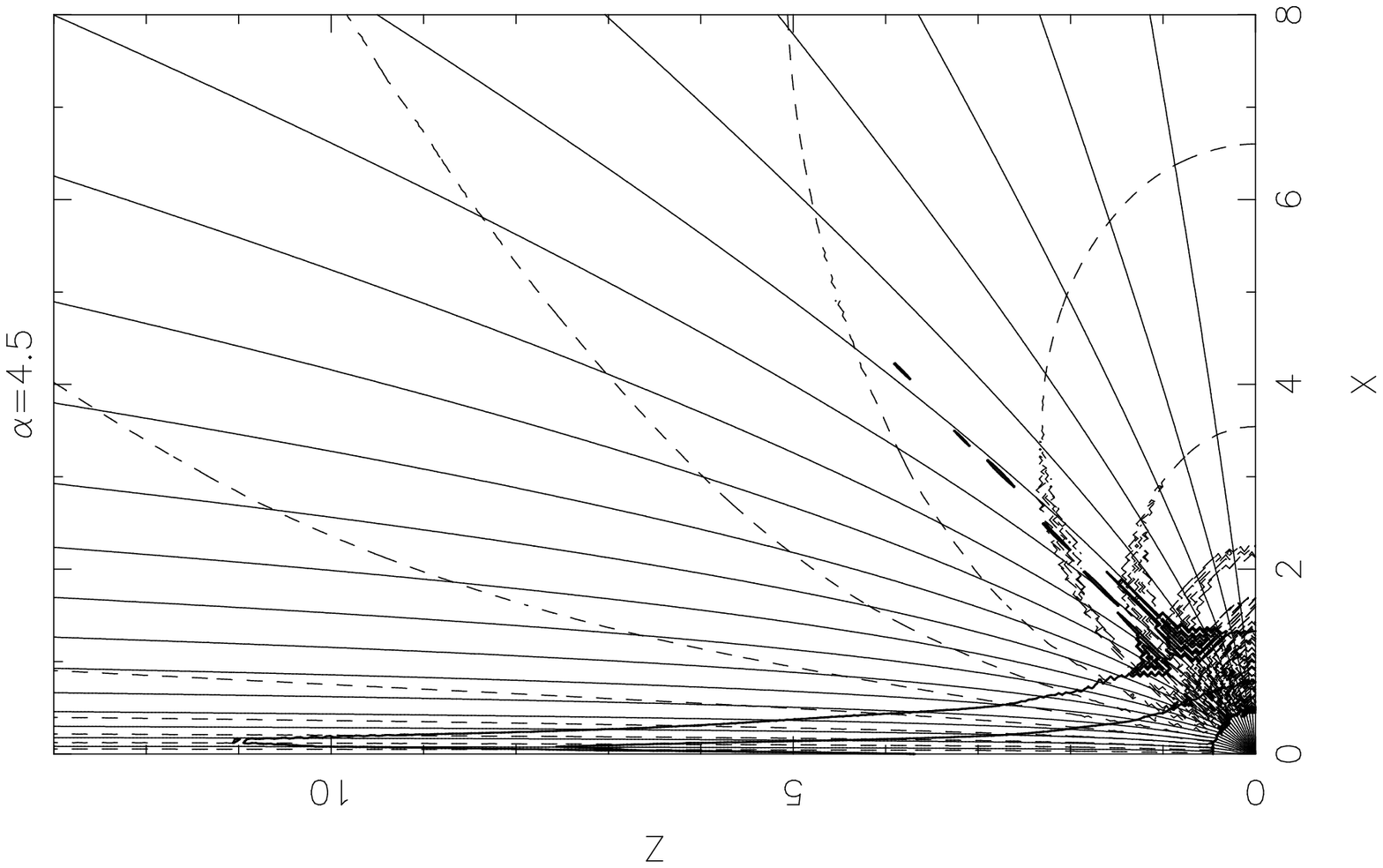,height=10.0truecm,angle=270}}
\setbox35=\vbox{\hsize=17 truecm 
{\noindent \footnotesize {\it Figure 3.}
Sequence of shapes of the poloidal field lines by increasing
the magnetic rotator parameter $\alpha$ from $\alpha=0.5$ (solar wind type
slow magnetic rotator) to $\alpha=4.5$ (fast magnetic rotator).
The initial nonrotating monopole magnetic field has a spherical Alfv\'en
surface located at $R=1$ and all 
distances are given in units of the Alfv\'en radius $R_a$ with the base 
located at $R=0.5$.
Dotted lines indicate poloidal currents. Thick lines
indicate Alfv\'en and fast critical surfaces. At $t=0$ the Alfv\'en surface
is spherical at $R=1$.
}}
\begin{figure*}
\centerline{\hspace{-3cm}\box31\hspace{-2cm}\box32}
\centerline{\hspace{-3cm}\box33\hspace{-2cm}\box34}
\vspace{1cm}
\centerline{\box35}
\end{figure*}

It follows that outflows from rapidly rotating stars and thin classical 
accretion disks may rather produce jets with characteristics which 
qualitatevely differ strongly from those of laminar jets usually discussed 
in the literature. This result is certainly obtained under the simplifying 
assumptions of the present study, and further investigations are necessary 
to explore the possibilities which may arise in this regime of outflow.
Up to now all numerical simulations avoided this problem by artificially
taking $\alpha \sim 1$ in order to obtain a stationary solution
(Romanova et al 1997), or, no stationary solution was obtained at all  
(Ouyed \& Pudritz 1997, see Fig. 8 in this work). 
Fig. (3)  then may indicate why no simulation has succeeded to
produce stationary supersonic jet from classical accretion disks up to
now (Ferreira, 1997).

\section{Results in the far zone}

The characteristics of magnetized outflows at large distances
from the central object have been studied by Heyvaerts \& Norman (1989),
Chiueh et. al. (1991) and Bogovalov (1995).
These studies have concluded that a stationary axisymmetrically rotating
object ejecting magnetized plasma always produces a jet collimated exactly 
along the axis of rotation, {\it if} the following conditions are satisfied:\\
1. The flow is nondissipative.\\
2. The angular velocity of rotation is nonzero everywhere (actually, if
$\Omega=0$ on some field lines, the same conclusion remains valid). \\
3. The total magnetic flux reaching infinity in any
hemisphere of the outflow is finite.\\
4. The polytropic index of the plasma  $\delta > 1$.\\
In such an outflow the density of the poloidal electric current is nonzero
in the region of the collimated flow. Conversely, in the region of
noncollimated field lines the density of the poloidal electric currents
equals to zero. This condition can be expressed by the constancy of the
quantity
$r^2\Omega (\psi)B_{p}/U_{p}$, which we shall call the
Heyvaerts-Norman integral.

\subsection{Efficiency of the magnetic rotator}

In the Weber \& Davis (1967) model of a magnetized equatorial wind the
terminal equatorial speed is  superfast with the fast mode surface placed
at some finate distance from the star. If the initial velocity of a cold plasma
is equal to zero on the surface of the star, the fast critical point is  
at infinity and we obtain the so-called Michel's (1969) minimum energy solution.
This solution is valid when the monopole-like poloidal magnetic field is
slightly disturbed by the plasma. In this case 
the asymptotic speed at infinity on the equator is $V_{\infty} = (3/2) V_a$,
while the {\it total} specific angular momentum in the system in units of
$R_a V_o$ is, $L = \Omega r_a^2/R_a V_o$. We recall that $R_a$ is the radius
of the initial spherical Alfv\'en surface while $r_a$ is the Alfvenic
radius on the equator at a given angular velocity.
An important physical quantity in magnetized outflows is the magnetic
rotator energy, $E_{MR}$, the product of the total specific angular momentum
and $\Omega$. The basal Poynting energy defined as the
ratio of the Poynting flux density $S_z$ per unit of mass flux density
$\rho V_z$ is approximately equal to $E_{\rm MR}$ {\it if} at the base of the
outflow the radius of the jet is much smaller than the Alfv\'en radius and
also the Alfv\'en number there is negligibly small.

The energy losses of the magnetic rotator per particle in units
of $V_o^2$ are
\begin{equation}
{E_{\rm MR}\over V_o^2}= {\Omega^2 r_a^2\over V_o^2} = \alpha L
\,.
\end{equation}
{ This formula is widely used in the theory of the rotational evolution
of stars. But the parameter $\alpha$ goes to $\infty$ in Michel's minimum 
energy solution. The plasma is strongly collimated under this condition and
therefore Michel's solution is not valid. 
In this case it is important to know how strongly
the effect of collimation affects those frequently used Michel's
energy losses.}
The energy losses $\alpha L$ per particle on the equator for Michel's
monopole-like solution and our calculations are compared in Fig. \ref{al}. 

\setcounter{figure}{3}
\begin{figure}
\centerline{\psfig{file=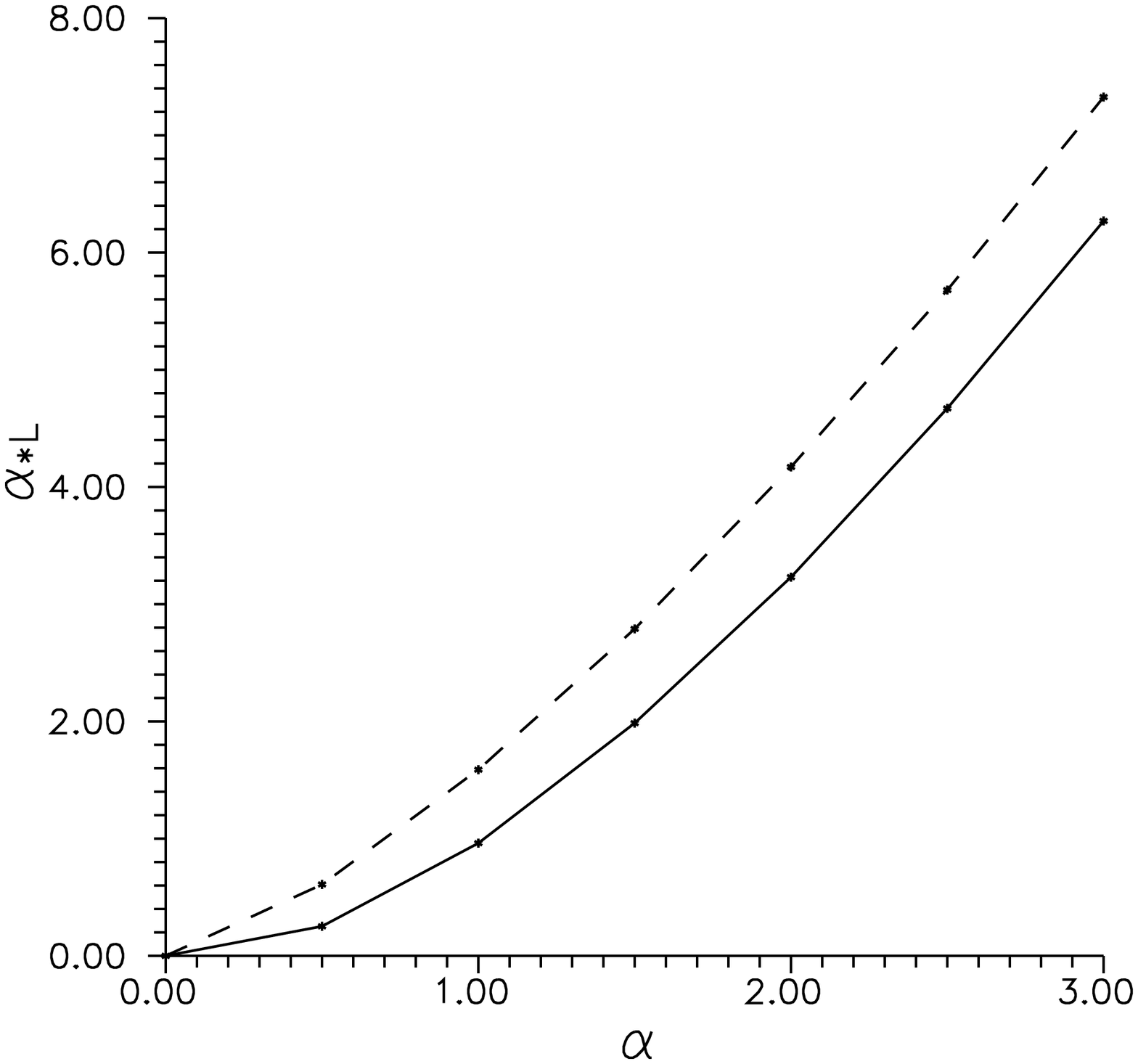,height=8.0truecm,angle=360}}
\caption{\it For a magnetic rotator we plot the rotational losses $\alpha L$ 
per particle, as a function of $\alpha$ (solid line). For comparison, the 
corresponding rotational losses for Michel's
minimum energy solution are also plotted (dashed line).
}
\label{al}
\end{figure}

The decceleration rate in our solution is less than in Michel's solution.
This is due to the collimation of the plasma.
The poloidal magnetic field near
the equator decreases and it leads to a reduction of the decceleration rate.

It is interesting that in spite of the decrease of the decceleration rate the
terminal velocity of the plasma near the equator increases in our solution.
The terminal speed $V_{\infty}$ as a function of $\alpha$ is plotted
in Fig. (\ref{term}). The terminal velocity is calculated on the equator at
the dimensionless distance X=500. For the case of a fast magnetic rotator
(Michel 1969),
the dependance of $V_{\infty}$ on $\alpha$ is,

$$
{V_{\infty}\over V_o}=
{1\over V_o}({\Omega^2 R_a^4 B^2_{a}\over \dot M})^{1/3}
= \alpha^{2/3}
\,,
$$
i.e., it goes like $\alpha^{2/3}$. This increase of the terminal velocity on
the equator in comparison to that in Michel's solution is due to the collimation
of the plasma. In the ollimated flow there exist strong gradients of
the toroidal magnetic field which additionally accelerate plasma. This means
that in collimated ouflows we have a more effective acceleration of the plasma by
the magnetic rotator.

\setcounter{figure}{4}
\begin{figure}
\centerline{\psfig{file=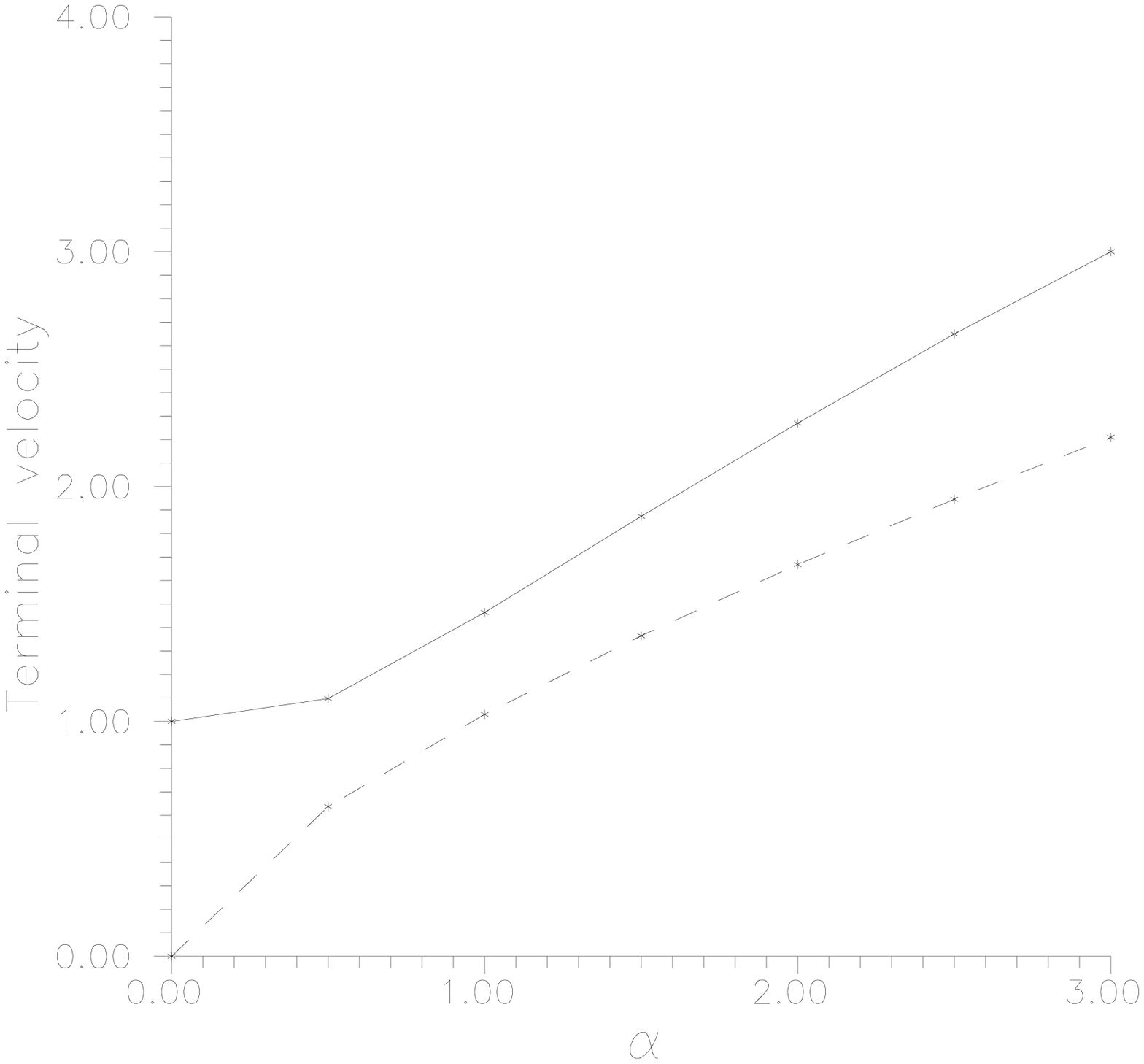,height=8.0truecm,angle=360}}
\caption{\it The terminal velocity ${V_{\infty}/ V_o}$ 
as a function of $\alpha$ (solid line).
For comparison, the corresponding terminal speed in Michel's minimum 
energy solution is also plotted (dashed line).
}
\label{term}
\end{figure}

The efficiency $e$ of the magnetic rotator in transforming part of the base
Poynting flux to poloidal kinetic energy at infinity is measured as the
difference of the poloidal kinetic energies at infinity and at the base
normalized to the total energy $E$ (in units of $V_o^2$).
The poloidal kinetic energy at infinity in Michel's solution in units
of $V_o^2$, $E^{pol}_{\infty}$ is
$$
E^{pol}_{\infty} = {E\over 3} = {\Omega^2 r_a^2\over 3 V_o^2}
\,,
$$
or,
$$
E^{pol}_{\infty} =
{1\over 3} {\Omega r_o\over V_o}^2 \left( {r_a\over r_o}\right)^2 =
{1\over 3} \alpha^2 \left( {r_a\over r_o}\right)^2 = {1\over 3} \alpha L
\,.
$$

The efficiency $e$ is plotted in Fig. (\ref{e}) for our solution and (solid)
and for Michel's solution (dashed). We see that due to collimation, 
the magnetic rotator becomes a very effective machine for plasma acceleration.

\setcounter{figure}{5}
\begin{figure}
\centerline{\psfig{file=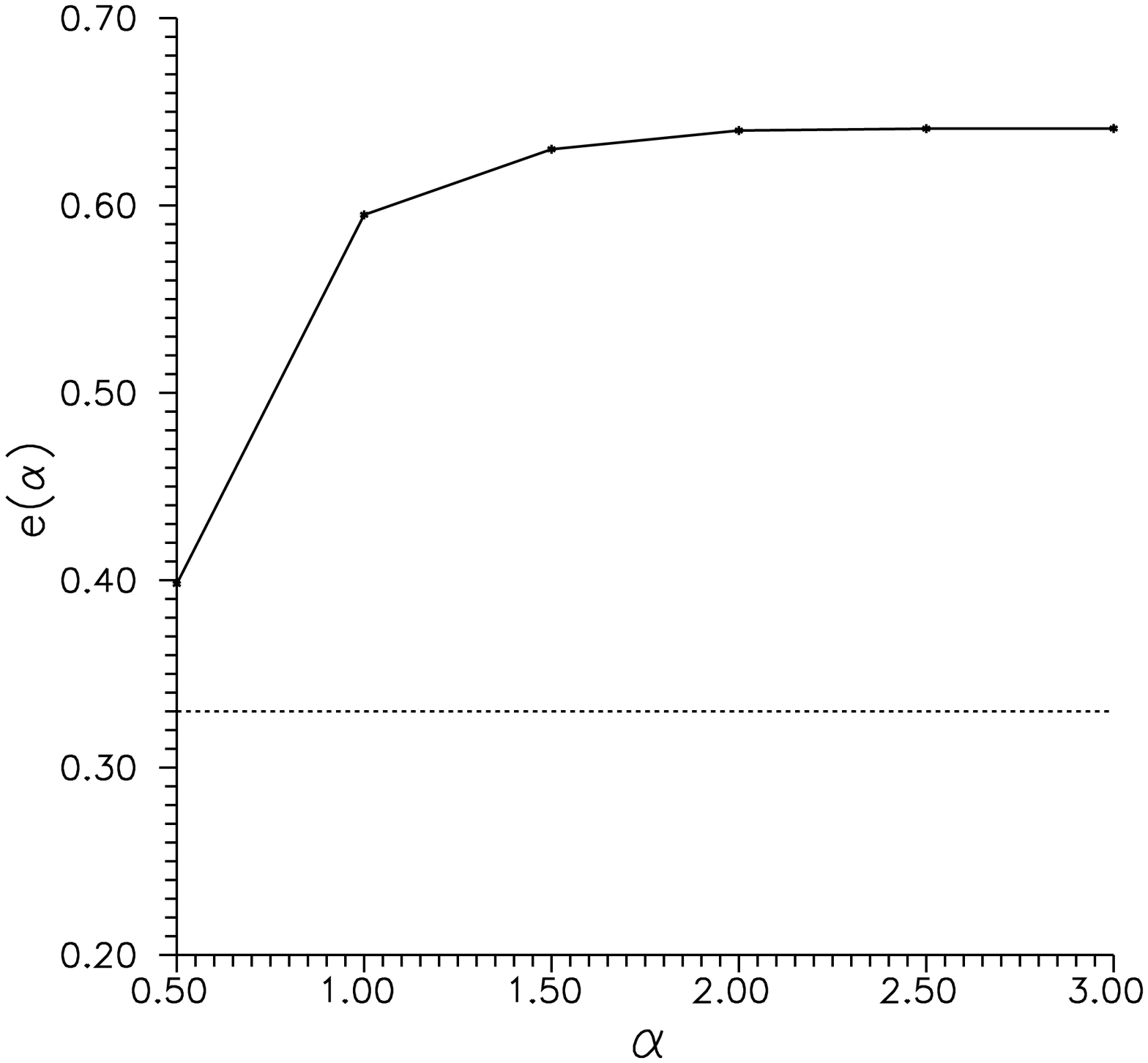,height=8.0truecm,angle=360}}
\caption{\it Efficiency $e$ of the magnetic rotator as a function
of $\alpha$ (solid line). For comparison, Michel's solution
has $e=1/3$ (dotted line).
}
\label{e}
\end{figure}

\subsection{The radius of the jet}

The dependence of the magnitude of the poloidal magnetic field and density
on the cylindrical distance $r$ becomes especially simple if we assume for
convenience that the following additional conditions are met in the
jet, namely that the integrals $W(\psi), L(\psi), F(\psi), \Omega(\psi)$ and
the terminal velocity $V_{j}$ are constants and do not depend on $\psi$
and also that $V_{j} \gg V_{a}(0)$, where
$V_{a}(0)$ is the Alfvenic velocity on the axis of rotation.
In such a case, Eqs. (\ref{trans2}) - (\ref{rpsi}) give an approximate
estimate of the dependence of the magnetic field on $r$ (Bogovalov 1995),

\begin{equation}
{B_{p}(r)\over B_{p}(0)}={\rho (r)\over \rho(0)}={1\over (1+(r/R_{j})^{2})},
\label{jr}
\end{equation}
where $R_{j}$  is the radius of the core of the jet
\begin{equation}
R_{j}=\sqrt{(1+{C_{s}(0)^{2}\over V_{a}(0)^{2}})}{\gamma V_{j}\over\Omega},
\end{equation}
with $C_{s}(0)$ the sound velocity along the jet's axis
and $B_{p}(0)$, $\rho(0)$
the magnetic field and the density of the plasma on the axis of the jet,
respectively.
It is evident that the poloidal magnetic field and the density remain
practically constant up to distances of order $R_j$ and then decay fast
like $1/r^2$ outside the jet's core.

Among the main goals of the present paper is the verification of these
conclusions, together with the investigation of the process of jet formation
and the development of methods for the calculation of the
characteristics of the plasma in the jet.

\setcounter{figure}{6}
\setbox77=\vbox{\hsize=7 truecm \vsize=7truecm
\psfig{figure=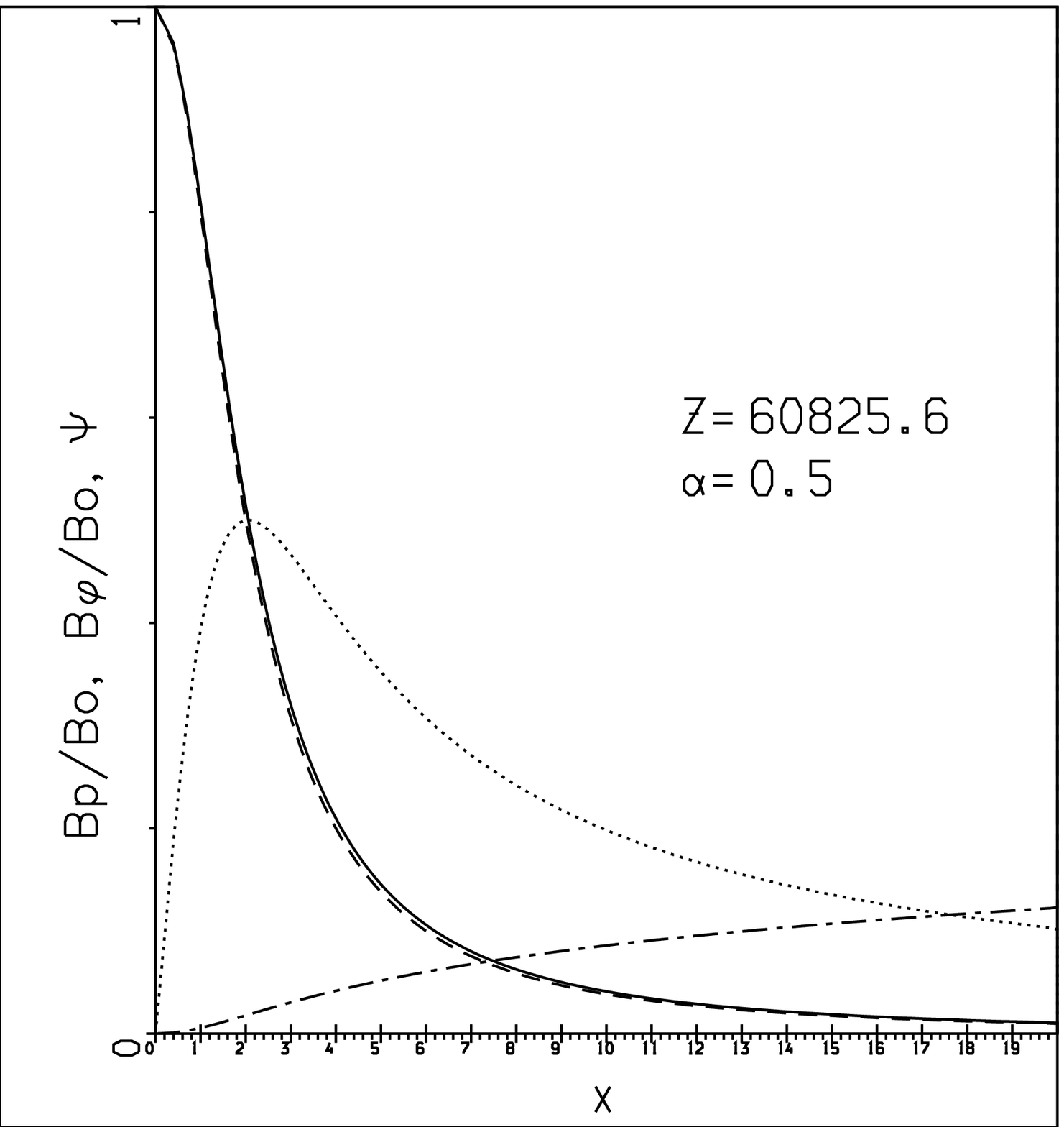,height=7.0truecm,angle=360}}
\setbox78=\vbox{\hsize=7 truecm \vsize=7truecm
\psfig{figure=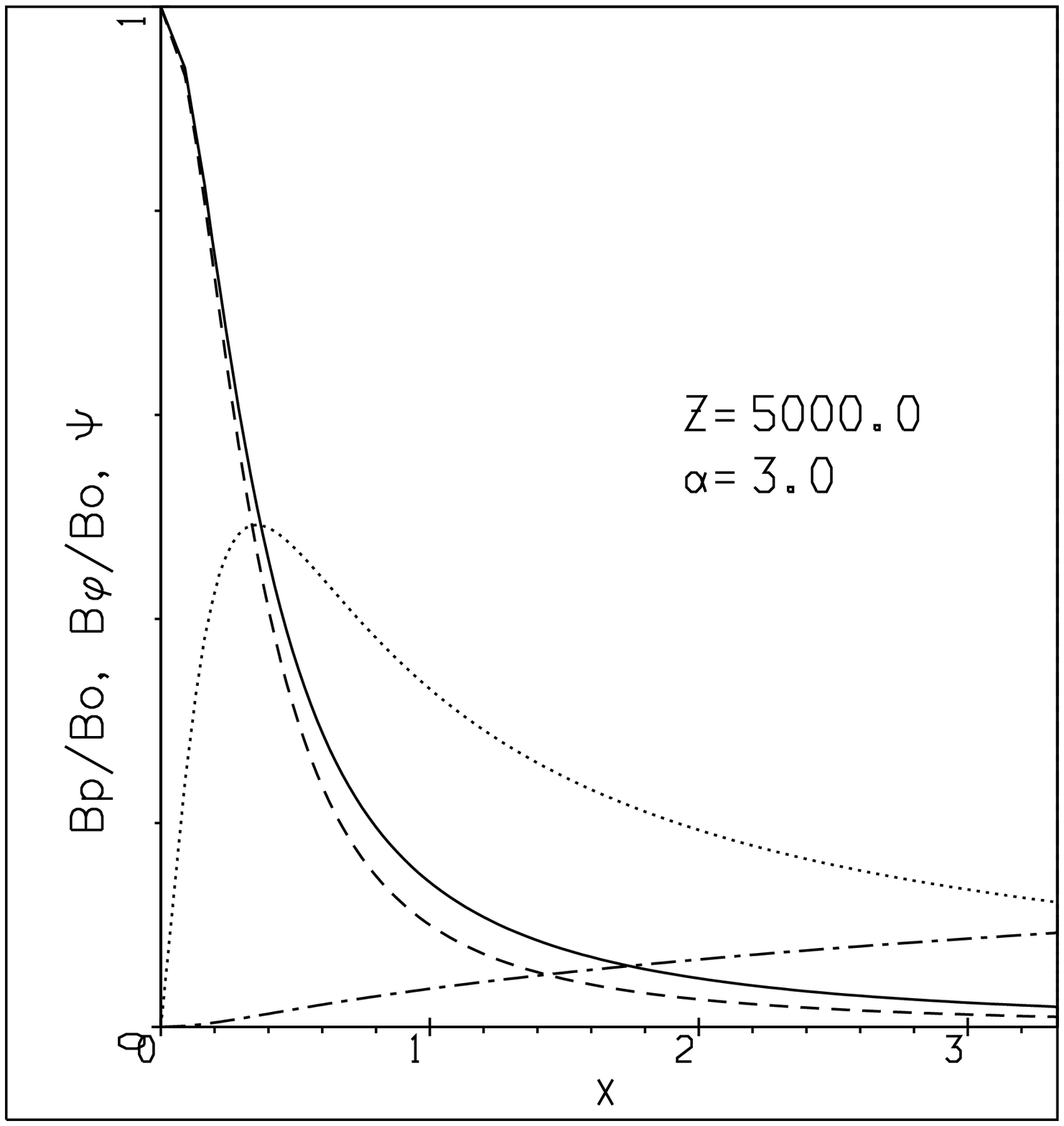,height=7.0truecm,angle=360}}
\begin{figure*}
\centerline{\box77\hspace{1.5cm}\box78}
\caption{\it Variation with the dimensionless cylindrical distance $X$
of the enclosed magnetic flux $\psi(X)$ (dot-dashes), strength of the
azimuthal magnetic field $B_{\phi}(X)/B_o$ (dots) and poloidal magnetic
field $B_p (X)/B_o$ (solid) for a slow magnetic rotator
$\alpha=0.5$ and a faster magnetic rotator $\alpha=3$. With dashes
the analytically predicted solution for the poloidal magnetic
field $B_p (r)/B_o$ is also shown.
}
\label{jets}
\end{figure*}

In Fig. (\ref{jets}) the poloidal and azimuthal components of the magnetic field
are plotted together with the magnetic flux enclosed by a cylindrical distance
$X$. The poloidal magnetic field $B_p (X)/B_o$
(solid line) is given in units of its reference value $B_o$,
corresponding to the magnetic field at the symmetry axis $r=0$
and some reference height $Z_o=60825$ for $\alpha=0.5$ and $Z_o=5000$
for $\alpha=3$. The asymptotic regime of the jet is achived at these
distances. The intensity of the poloidal
magnetic field drops dramatically by more than 50\% with respect to its
value at the axis $X=0$ within a distance of the order of the
collimation radius $X_j=1/\alpha$. The dashed curve gives
the analytically predicted solution for the poloidal magnetic
field $B_p (X)/B_o$, Eq. (\ref{jr}). Apparently, for small values of $\alpha$
there is a good agreement between the calculated and analytically
predicted values of the poloidal magnetic field. For $\alpha=3$ the
agreement between the analytical prediction and numerical calculations is
worse than for $\alpha=0.5$. This is natural because the condition
$V_{j} \gg V_{a}(0)$ under which the analytical prediction is valid
becomes not so good fulfilled for $\alpha=3$ as for $\alpha=0.5$.

The strength of the azimuthal magnetic field $B_{\phi}(X)/B_o$ (dots)
obtains a maximum value at the radius $X_j$.
For $X < X_{j}$ we have approximately conditions similar to those
corresponding to a uniform current density wire and therefore $B_{\phi}
\propto X$. On the other hand, for $X > X_{j}$, conditions like those
existing outside of a uniform current density wire exist and therefore
$B_{\phi} \propto X^{-1}$.

\setcounter{figure}{7}
\setbox101=\vbox{\hsize=8 truecm \vsize=7truecm
\psfig{figure=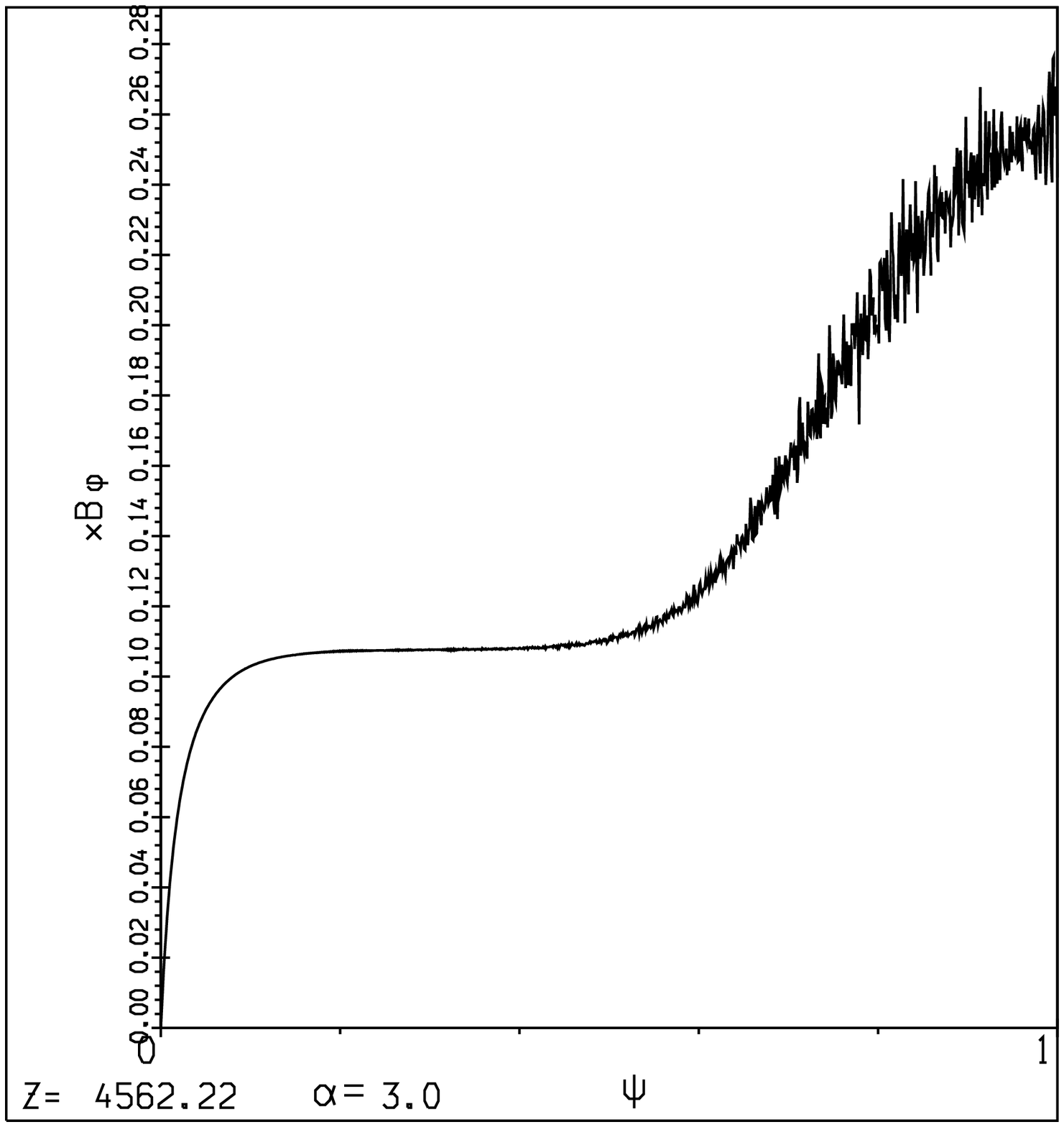,height=8.0truecm,angle=360}}
\setbox102=\vbox{\hsize=8 truecm \vsize=8truecm
\psfig{figure=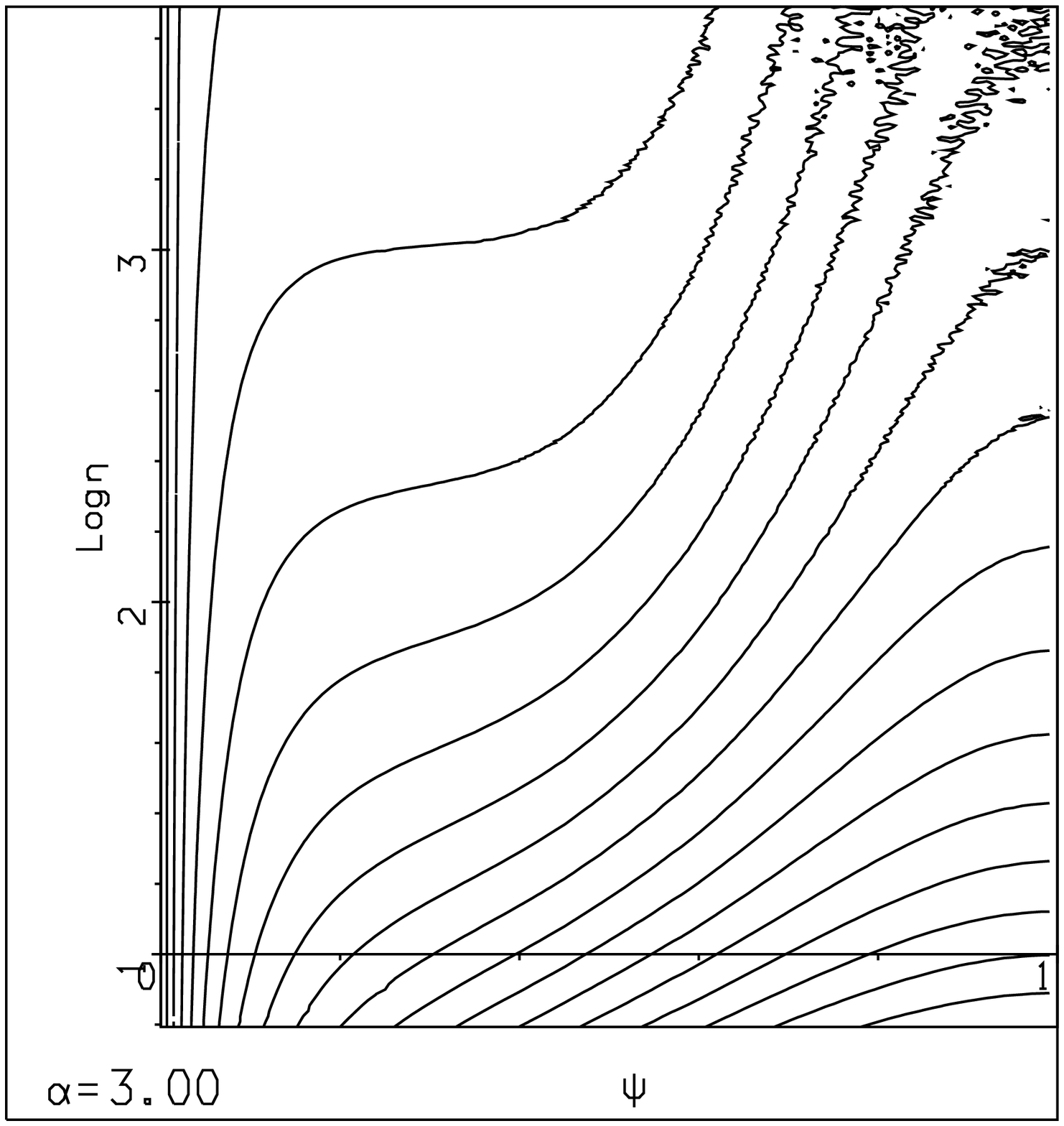,height=8.0truecm,angle=360}}
\begin{figure*}
\centerline{\box101\hspace{1.5cm}\box102}
\vspace{1cm}
\caption{\it
For $\alpha =3$, in (a) it is plotted the enclosed
electric current $XB_{\phi}$ by each poloidal field line $\psi$ and
in (b) the iso-current contours in the space of $(\eta, \psi)$.
}
\label{current}
\end{figure*}

\subsection{The Heyvaerts \& Norman enclosed current}

The enclosed electric current by a poloidal field line $\psi =const.$
is plotted in Fig. (\ref{current}a) for $\alpha=3$. We may distinguish three
regimes. Close to the axis, $0< \psi < 0.1$ the current
increases with $\psi$ since there we have conditions similar to those
corresponding to some uniform current density wire. A new regime appears
when we are outside the collimated region where the enclosed current
reaches a plateau, $0.1 < \psi < 0.5$. We shall call this regime the
Heyvaerts-Norman regime, since there we have conditions similar to those
existing outside a uniform current density wire ($XB_{\phi} = const.$),
a situation described by Heyvaerts \& Norman (1989). Finally, a third
domain exists in $0.5< \psi < 1$ where again the enclosed current
increases. It seems that this contradicts the expected behaviour, but
as we show below it is due to very slow decrease of this part of the
current and we do not achieve final asymptotics.

\subsection{Logarithmic collimation asymptotically}

\setcounter{figure}{8}
\setbox1=\vbox{\hsize=7 truecm \vsize=7truecm
\psfig{figure=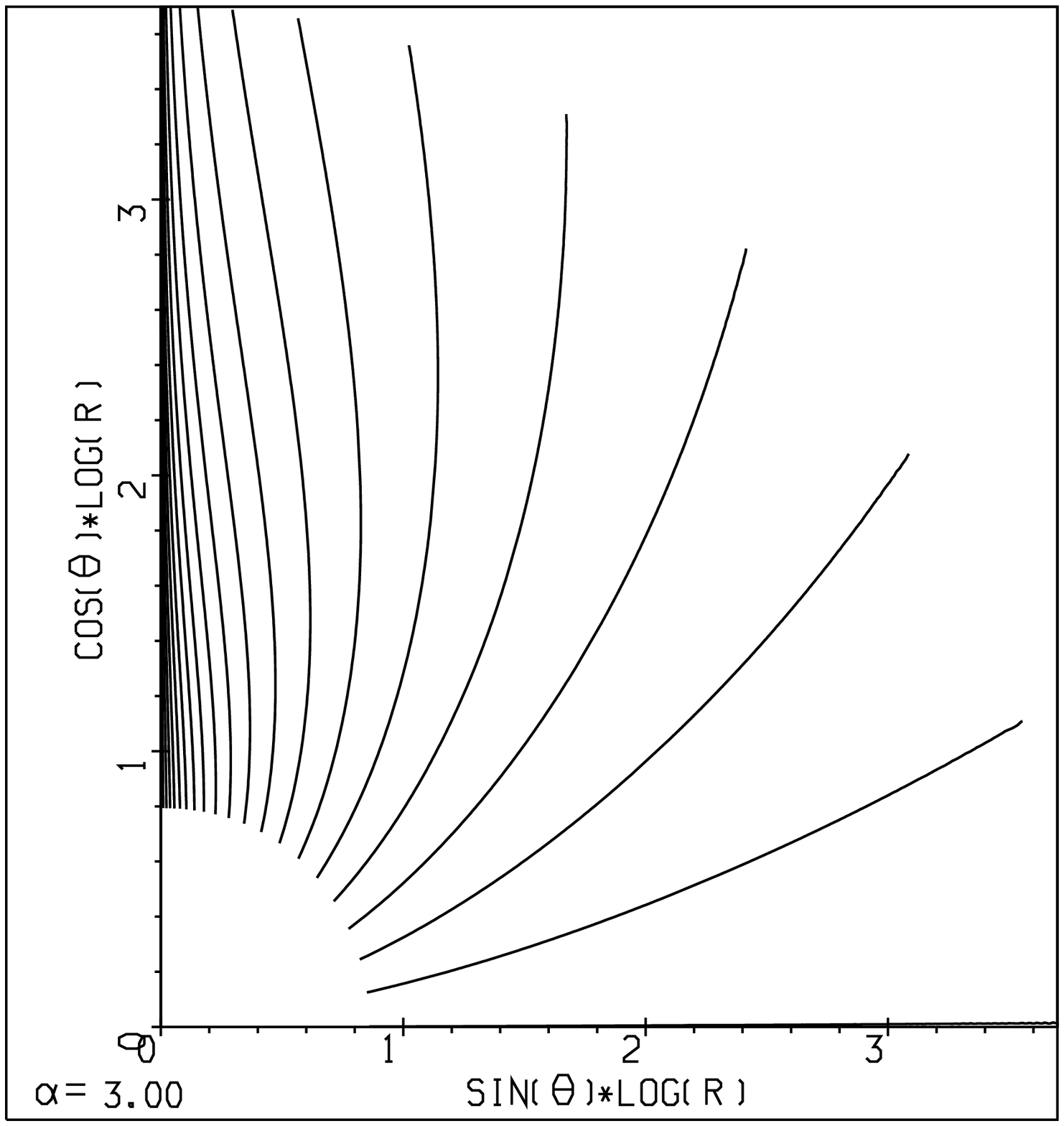,height=7.0truecm,angle=360}}
\setbox2=\vbox{\hsize=7 truecm \vsize=7truecm
\psfig{figure=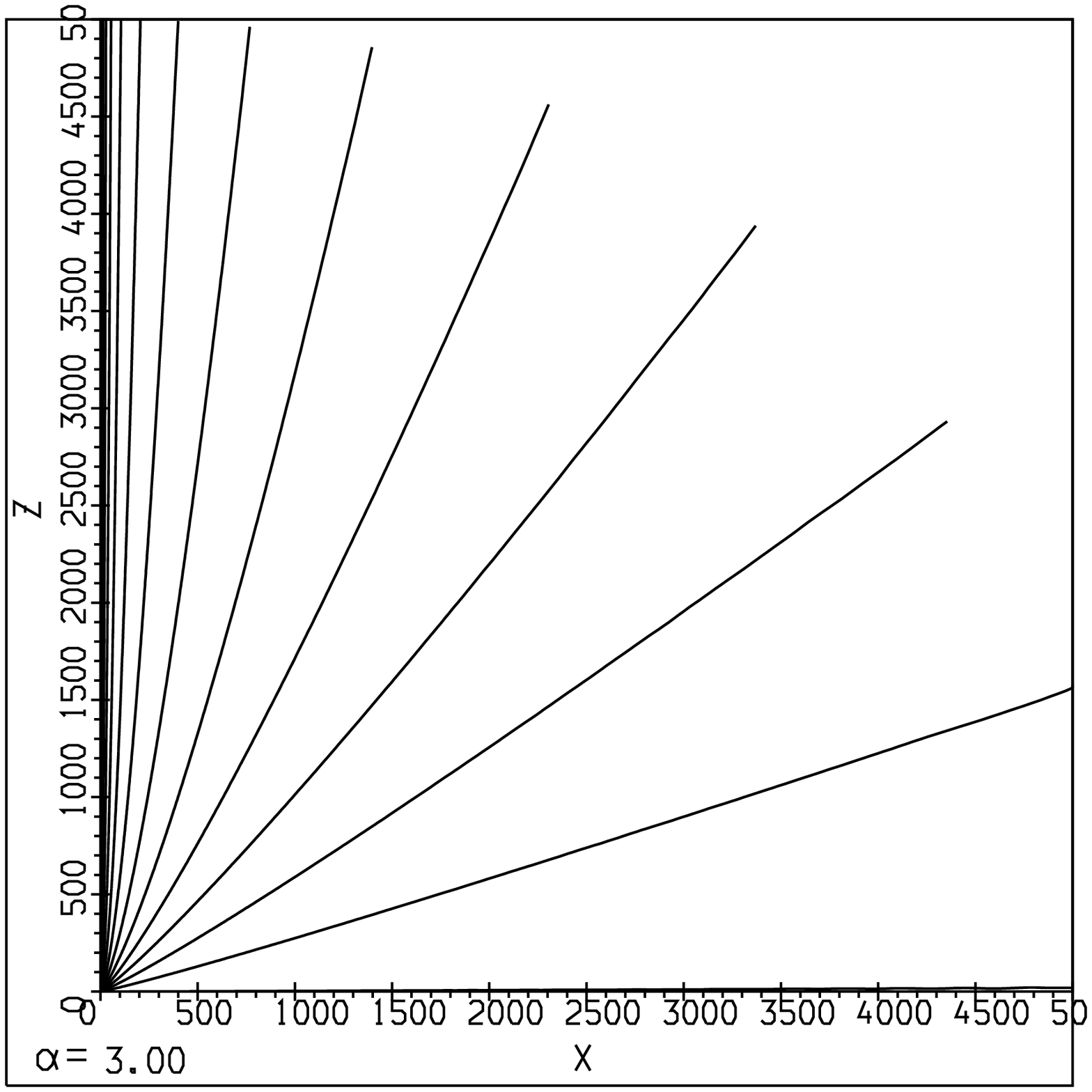,height=7.0truecm,angle=360}}
\begin{figure*}
\centerline{\box1\hspace{1.5cm}\box2}
\vspace{1cm}
\caption{\it
Shape of the poloidal magnetic field lines for a
magnetic rotator parameter $\alpha=3$ plotted up to a radius 5000 times
the initial base radius.  In (a) the poloidal field lines are plotted
in a logarithmic scale, and give the erroneous impression that all 
fieldlines are
focused towards the system's symmetry axis.  In the linear scale of
Fig. (b) however, a cylindrical jet is formed around the symmetry axis
while all other field lines go to straight asymptotes which fill all space.
}
\label{flow}
\end{figure*}

One may use the coordinates  $(\eta, \psi)$ to show the distribution
of the electric currents, Figs. (\ref{current}b).
In this space, $\psi =0$ corresponds
to the symmetry axis $X=0$,  $\psi = 1$ to the equator and 
$\eta =0$ approximately to the source surface,
while a poloidal fieldline corresponds to some vertical line $\psi =const.$
Fig (\ref{current}b) demonstrates that there is an electric current near 
the axis. Then the
Heyvaerts \& Norman region of constant $XB_{\varphi}$ is formed. And finally
the region where $XB_{\varphi}$ increases is observed. It is important
that the excess of the electric current in this last region decreases with
$\eta$. But this decrease occurs in a logarithmic scale. The solution goes
to its asymptotic form, but this is done very slowly,  in a logarithmic scale.
This behavior can be demonstrated also by a simple analysis of the transfield
equation.

In a supersonic flow, the motion of every parcel of plasma
is controlled by the initial conditions and the forces affecting the plasma.
Here we are interested in the collimation process in the region of
radially expanding field lines. In this region forces due to the
poloidal magnetic
field and the inertia of the plasma due to motion in the azimuthal direction
can be neglected in the transfield equation in comparison to forces arising
from the
toroidal magnetic field. The terms $B_{p}^{2}/8\pi$ and
$\rho V_{\varphi}^2$ can be neglected since they drop with
distance r as $1/r^{4}$, while the
terms corresponding to the azimuthal magnetic
field $B^2_{\varphi}$ drop as $1/r^{2}$.
Then, the transfield equation Eq. (\ref{trans2}) is simplified as follows,

\begin{equation}
{1\over 8\pi r^{2}}{\partial\over
\partial \psi}(rB_{\varphi})^{2}
-{U_{p}\over 4\pi rR_{c}F(\psi)}=0.
\label{trans3}
\end{equation}

It may be seen from this equation that the curvature radius of a field
line of the poloidal magnetic field is defined by 
the tension of the toroidal magnetic field. Let us estimate  how the
curvature radius of the poloidal magnetic field changes with distance assuming
for simplisity that the poloidal magnetic field expands radially and hardly
depends on the polar angle $\theta$ near equator. At large distances,
the frozen in condition gives for the electric current

\begin{equation}
rB_{\varphi}=-{r^{2}\Omega B_{p}\over V_{p}}.
\label{freez2}
\end{equation}
Inserting this in (\ref{trans3}) and assuming that the dependence of
$B_{p}$ on $\theta$ is weak, we get

\begin{equation}
{2\cos{\theta}B_{p}\Omega^{2}\over V_{p}^{2}}={V_{p}\over rR_{c}Fc}.
\end{equation}

The curvature radius of the field line is defned by the equation
\begin{equation}
R_{c}={dl\over d\theta},
\end{equation}
where $dl$ is the element of the length of the fied line. Since
to a first approximation $dl=dR$, where $R$ is the spherical distance to the
center of the rotating object, we get for the equation of a field line,

\begin{equation}
{d\theta\over dR}={cFB_{p}R^{2}\Omega^{2}\over RV_{p}^{3}} \sin{2\theta}
\,.
\label{line}
\end{equation}
For a radially expanding magnetic field $B_{p}R^{2}$ is constant along a
magnetic field line. Therefore, the turn angle of the field line
$\Delta\theta$ depends logarithmically on $R$,

\begin{equation}
\Delta\theta={ cFB_{p}(R\Omega)^{2} \sin{2\theta}
\over V_{p}^{3}}\ln{{R\over R_{0}}},
\label{teta}
\end{equation}
where $R_{0}$  is the initial distance. According to (\ref{freez2}) the change of the electric current enclosed by
a field line is connected with the divergence of the poloidal magnetic
field from the purely radial shape.
To have variation of $rB_{\varphi}$ with $\psi$
it is necessary to turn some field
lines at some angle.   Such a turn occurs at the exponentially large distances
defined by the expression
\begin{equation}
R=R_{0}\exp{ ({ V_{p}^{3}\Delta\theta\over
cFB_{p}(R\Omega)^{2} \sin{2\theta} } )} 
> R_{0}\exp{\Delta\theta\tan\theta \over 2 }.
\label{dist}
\end{equation}
because for the super fast magnetosonic plasma we have
\begin{equation}
V_{p} > cFB_{\varphi}^{2}/B_{p},
\end{equation}
which taking into account (\ref{freez2}) can be rewritten as
$V_{p}^{3} > cFB_{p}r^{2}\Omega^2$.
In other words, the magnitude of the exponent in Eq. (\ref{dist}) is larger
than $\Delta\theta\tan\theta /2$, a large number near the equator.
Apparently, two points at small angular distance $\Delta \theta$ on
the same fieldline have exponentially large radial distances $R_{0}$ and
$R$ from the origin. This explains the picture we obtain in the numerical
solution and agrees with earlier results obtained by Eichler (1993) and
by Tomimatsu (1994).

In the logarithmic scale of Figs. (\ref{flow}a) the impression one may get
is that \underbar{all} poloidal fieldlines are focused towards the
symmetry axis. Such a figure was produced in Sakurai (1985).
However, this way of plotting the shape of the poloidal streamlines is
a deceiving about the asymptotical shape of the magnetosphere which
is different, as seen  in the linear scale of Fig. (\ref{flow}b).
In this plotting it is clear that the geometry of the poloidal field
lines is such that a cylindrical core (the jet) is formed around the
symmetry axis with a width of the order of the base radius, but all
other field lines go to straight asymptotes filling all space.

\section{Accretion disk-like Rotation, $\Omega =\Omega(\psi )$.}

An analysis of the asymptotical behaviour of MHD outflows in 
Bogovalov (1995) shows that a discontinuity in the total electric current
is formed on the equator for outflows from astrophysical objects having 
a magnetic field directed in opposite directions in the upper and 
lower hemispheres.  
In a pure monopole-like magnetic field the electric current leaves the star 
in the upper hemisphere and returns back in the lower hemisphere. 
In such a magnetosphere the total electric current is continuous on the 
equator. But already in the split monopole model where
the magnetic field has opposite directions in the upper and lower hemispheres,
the closure of the electric circuit occurs along the equator so that on the
equator $XB_{\phi}=0$. This means that the total electric current flowing
in the upper or lower hemispheres is equal to zero.
On the other hand, according to Bogovalov (1995) the magnitude of $XB_{\phi}$
is not equal to zero on all field lines of the supersonic flow where
$\Omega \ne 0$. This means that the function
$XB_{\phi}(\psi)$ shown in Fig. (\ref{current}a) has actually a discontinuity
at the point $\psi =1$ (on the equator) for the split monopole solution.
This is the usual MHD contact discontinuity between the flows in the 
upper and lower hemispheres with the magnetic field equal to each other in 
magnitude but with different directions. It is important to check the validity
of this conclusion for a pure monopole-like magnetic field.

In the following, we shall investigate the process of the formation on the 
equator of such a discontinuity in the total electric current, in the ideal MHD 
approximation.
For this purpose, we take the rotation of the star such that 
the total electric current which flows in each of the upper or lower
hemispheres to be independently equal to zero. For example, this happens 
with a differential law of rotation wherein $\Omega(\psi)=0$ on the equator 
and the fieldline on the equator does not rotate.
This guarantees that $XB_{\phi}=0$, everywhere on the equator.

\setcounter{figure}{9}
\begin{figure}
\centerline{\psfig{file=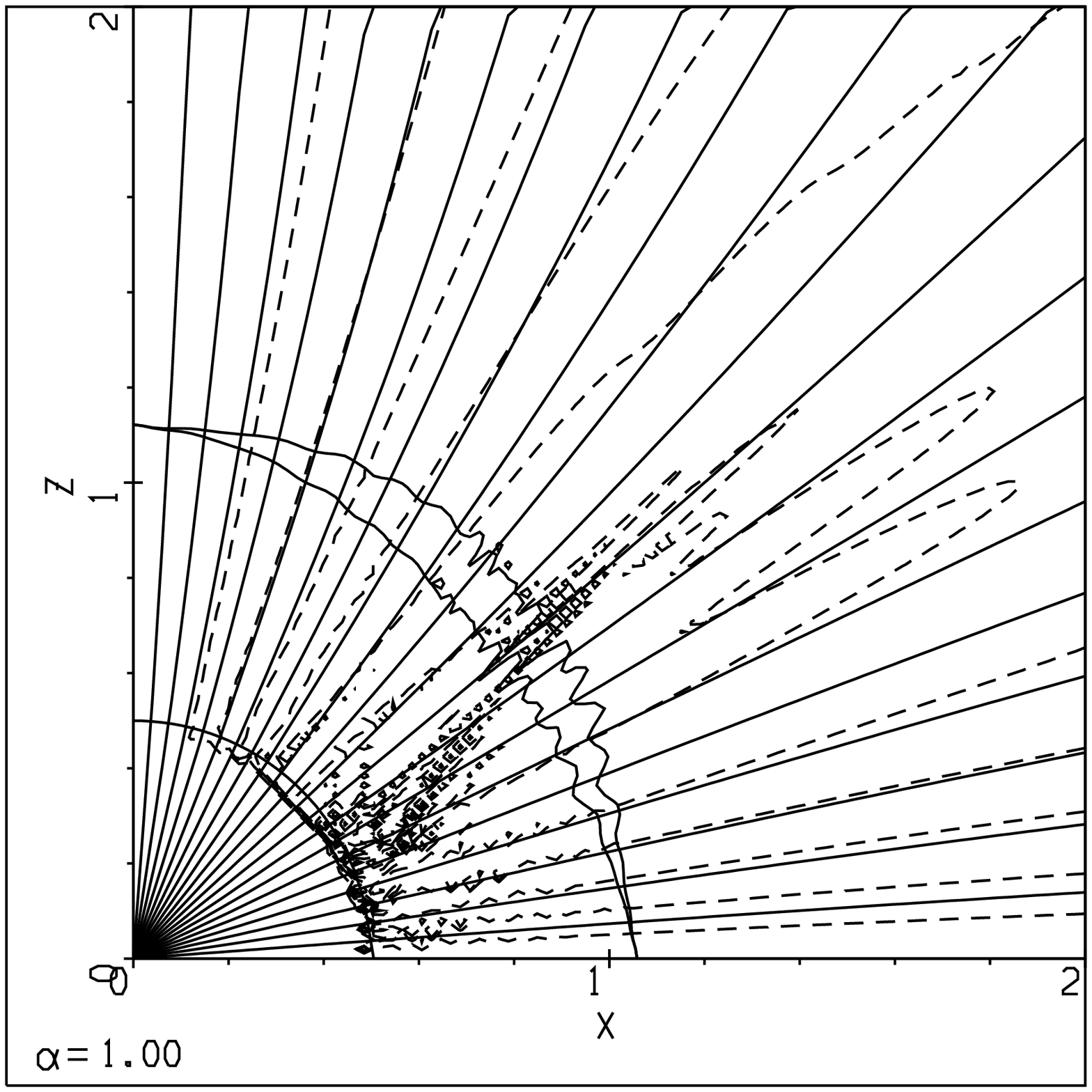,height=8.0truecm,angle=360}}
\caption{\it Shape of the poloidal magnetic field lines in the near
zone of a differentially rotating magnetic rotator with $\alpha=1$: 
the field lines focus towards the pole {\it and} the equator.
}
\label{disk1}
\end{figure}

\setcounter{figure}{10}
\begin{figure}
\centerline{\psfig{file=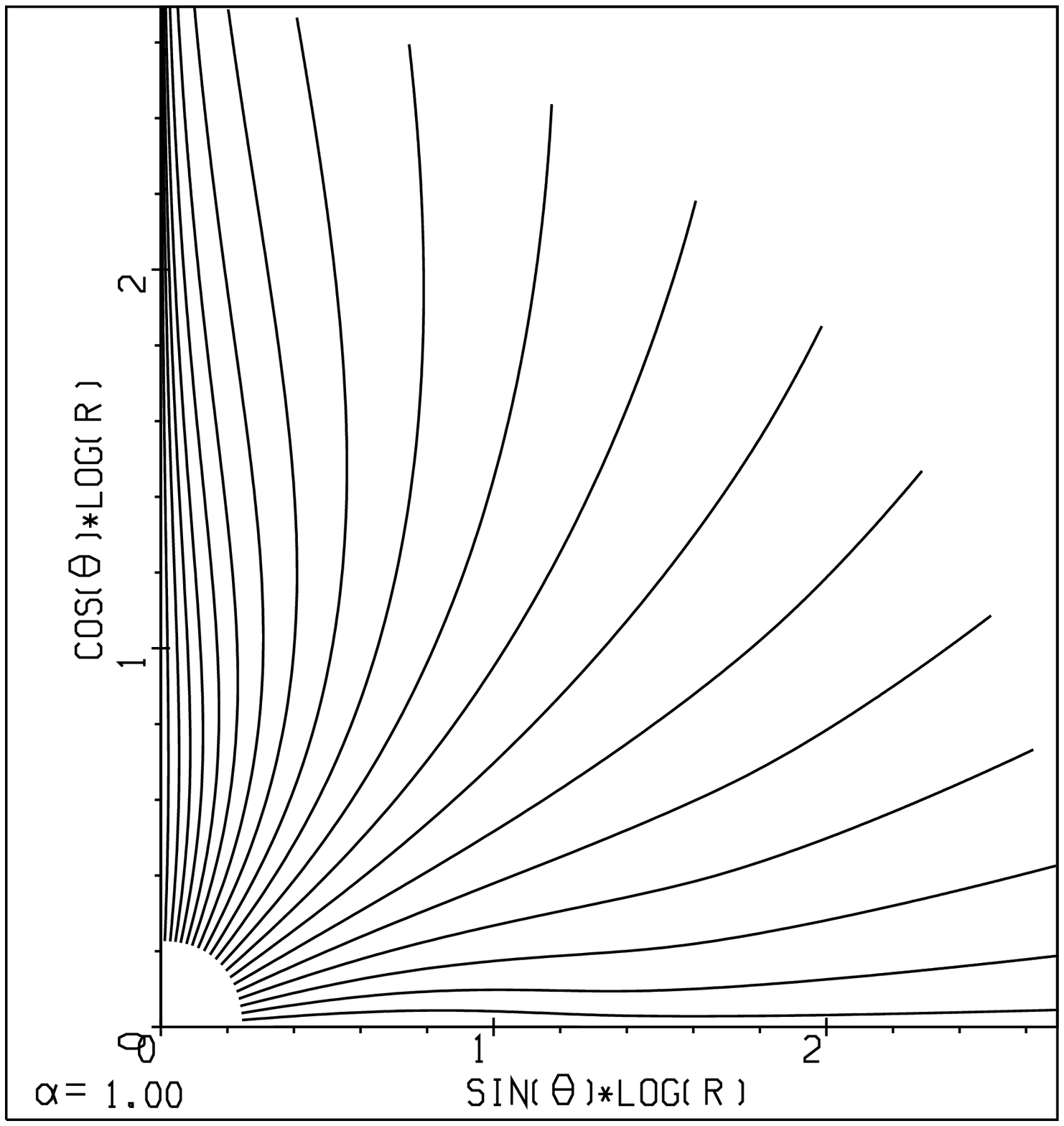,height=8.0truecm,angle=360}}
\caption{\it Shape of the poloidal magnetic field lines for a
differentially rotating magnetic rotator with $\alpha=1$, 
plotted in a logarithmic scale
up to a radius 1000 times the initial base radius.
}
\label{disk2}
\end{figure}

In this work we took the simplest law of rotation
\begin{equation}
\Omega(\psi)=\alpha  (1-\psi)\,.
\end{equation}
It is worth to pay attention to the fact that that this law of rotation 
describes qualitatively the rotation of accretion disks where the largest
angular velocity is in the inner edge of the disk and the lower one at the
outer edge of the disk. The solution of the problem for
this differential law of rotation in the nearest zone is shown in Fig.
(\ref{disk1}). 
\setcounter{figure}{11}
\setbox201=\vbox{\hsize=8 truecm \vsize=7truecm
\psfig{figure=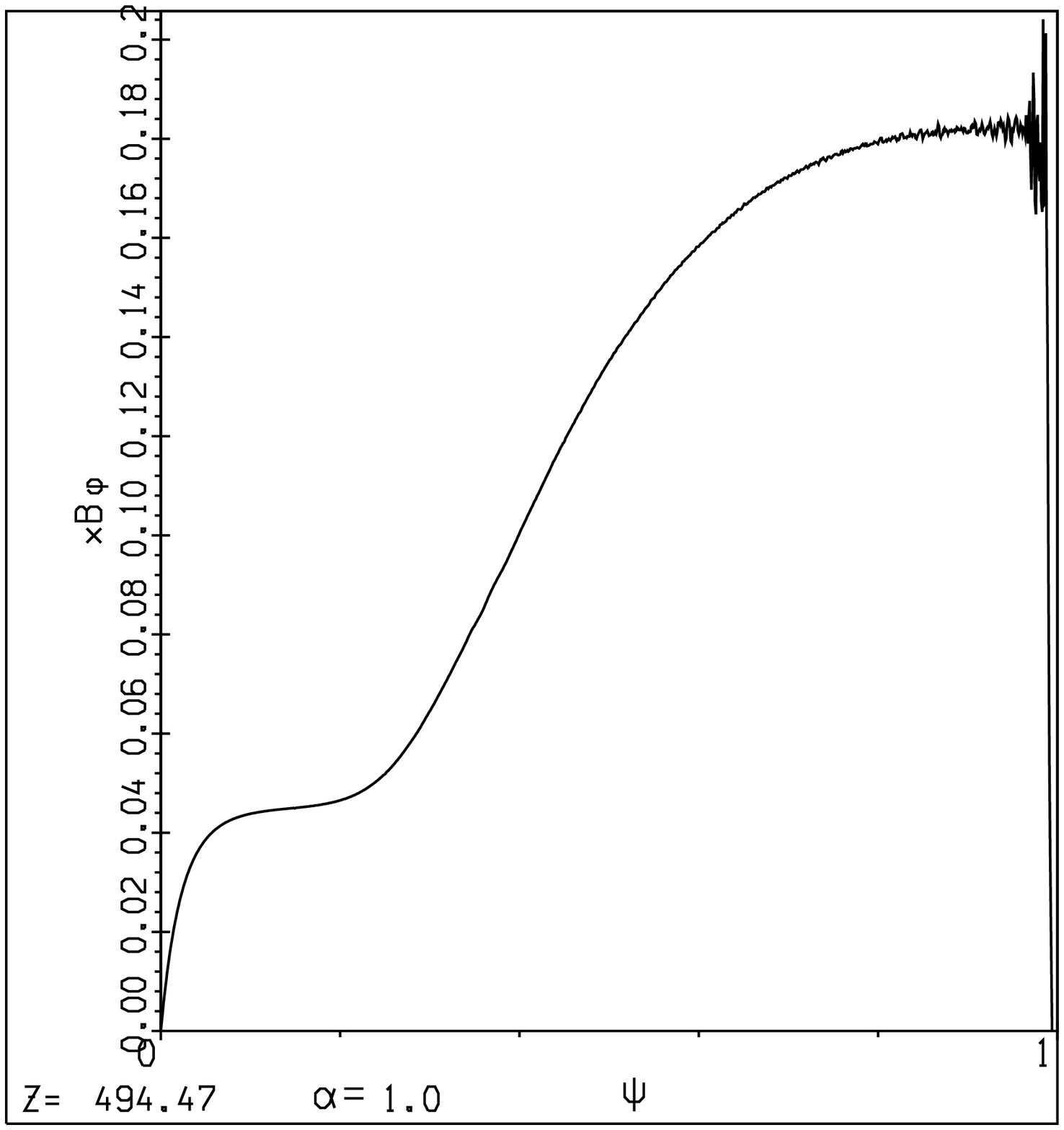,height=8.0truecm,angle=360}}
\setbox202=\vbox{\hsize=8 truecm \vsize=8truecm
\psfig{figure=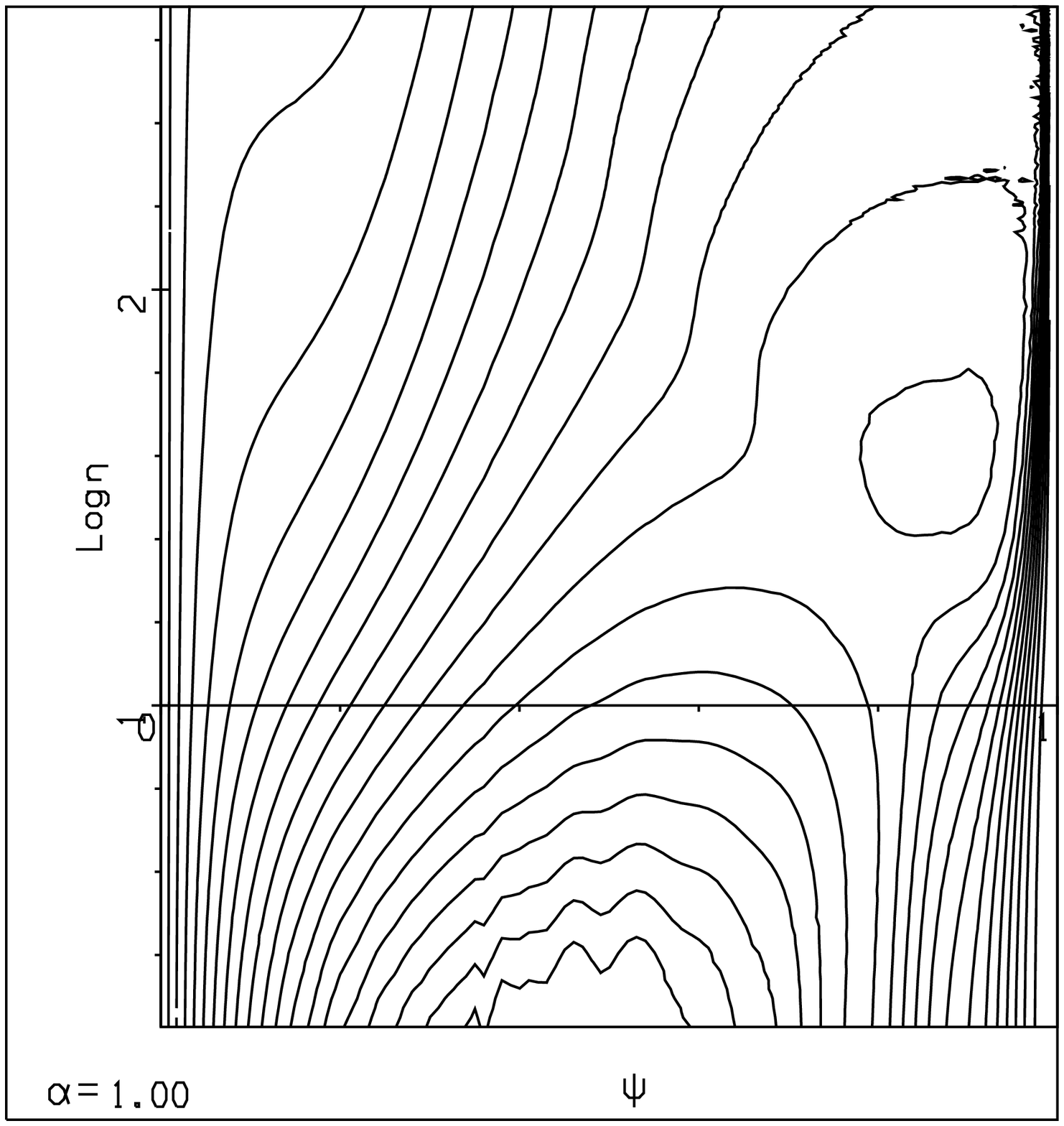,height=8.0truecm,angle=360}}
\begin{figure*}
\centerline{\box201\hspace{1.5cm}\box202}
\vspace{1cm}
\caption{\it
For a differentially rotating magnetic rotator with $\alpha =1$, 
in (a) it is plotted the enclosed
electric current $XB_{\phi}$ by each poloidal field line $\psi$ and
in (b) iso-current contours in the space of $(\psi , \eta )$.
}
\label{dcur}
\end{figure*}

The most important difference with isorotation is the coincidence
of the fast mode and Alfvenic surfaces on the equator. This is simply the 
consequence of the assumed zero toroidal magnetic field on the equator.

The poloidal magnetic field in the far zone is shown in a logarithmic scale in
the next Fig. \ref{disk2}. It is clearly seen in this figure that along
with the expected collimation of the flow towards the axis of rotation, there is
also a focusing of the flow towards the equator. 
The explanation of this
result is simple. With the assumed differential law of rotation, the 
largest toroidal magnetic field in the nearest zone is found somewhere 
at the middle latitudes.  The pressure of this toroidal magnetic field 
pushes the plasma towards the equator and the pole from those mid latitudes.

Fig. (\ref{dcur}) shows the formation of a discontinuity in the total
electric current near the equator ($\psi = 1 )$. 
On the other hand, iso-current contours in Fig. (\ref{dcur}b) show that at 
small $\eta$ the dependence of $XB_{\phi}$ on $\psi$ is that of a smooth 
function with a maximum placed at $\psi =0.5$. This maximum moves to the
equator as $\eta$ increases and finally the distribution of  $XB_{\phi}$
is similar to the one for an isorotation as that shown in Fig. \ref{current},  
with the exception of a discontinuity formed near the equator.

\section{Relativistic outflows}

In this section we discuss briefly the problem of collimation of a
relativistic plasma. This relativistic version of the problem under
consideration differs from the nonrelativistic one only in the modification
of the function $G(\eta, \psi)$. In the relativistic case, the terms with
the electric field play an important role. To specify the
boundary conditions in the
super fast magnetosonic region we will use the approximate solution obtained
in Bogovalov (1997) for a relativistic plasma outflow in the nearest
zone. In this solution the magnetic flux function is

\begin{equation}
\psi= 1-cos{\theta}
\,,
\end{equation}
corresponding to the poloidal magnetic field of a magnetic monopole and

\begin{equation}
B_{\varphi}=-xB_{p}{\gamma_{0}\over U_{0}}\,,
\end{equation}
where $\gamma_{0}$ is the initial Lorentz factor of the plasma which is
ejected from the surface of the star, $U_{0}$ is the initial four-velocity
of the outflow, $x$ is the cylindrical distance in units of the
light cylinder radius. We assume a uniform rotation of the star, 
\begin{equation}
U_{p}=U_{0}, \qquad
U_{\varphi}=0
\,, \qquad
n(r)={n_{0}\over r^{2}}\,,
\end{equation}
where $n_{0}$ is the density of the plasma on the light cylinder. The solution
written above is an approximate one, with corrections in the subfast
magnetosonic region of the order $\sigma/\gamma_{0}^{3}$, where
$\sigma=(B_{0}^{2}/ 4\pi mc^{2}n)({R_{star}\Omega /c})^{2}$
is the Poynting flux per particle at the equator.
The fast mode surface in these variables is
$r_{f}=\sqrt{{\sigma/\gamma_{0}}}$. This solution is valid under
the condition $\sigma/\gamma^{3} \ll 1$. To get an idea of the physical
parameters it may be useful to recall that for the Crab pulsar we have
$\sigma \sim 10^{6}$,
$\gamma_{0}\sim 10^{3}$, and
$\sigma/\gamma_{0}^{3}\sim 10^{-3}$.

The boundary conditions in the form of the integrals
$W(\psi)$, $L(\psi)$, $\Omega(\psi)$,  $F(\psi)$ were specified directly after
the fast mode surface and the solution obtained is shown in Fig. (13)
for sigma $\sigma = 300$ and $\gamma_{0}=30$.
The collimation of the flow in this relativistic case is very weak. To
explain this let us estimate the dependance of the radius of curvature of the
star on the distance to the star assuming that the magnetic field to a
first approximation is the field of the magnetic monopole. At large
distances,  we can neglect the poloidal magnetic field and the
azimuthal rotation of the plasma.  In this case the transfield equation
becomes,

\begin{equation}
{1\over 8\pi x^{2}}{\partial\over
\partial \psi}(x^{2}(H^{2}_{\varphi}-E^{2}))
-{(U_{p}+F(\psi)x^{2}B_{p})\over 4\pi xR_{c}F(\psi)}=0.
\label{trans4}
\end{equation}
It is easy to obtain similarly to the nonrealtivistic case the approximate
equation for a field line,

\begin{equation}
{d\theta\over dr}={\sin{2\theta}\over rU_{0}^{2}({U_{0}\over\sigma}
+\sin^{2}{\theta})}.
\label{eq1}
\end{equation}
An integration of this equation gives approximately,

\begin{equation}
r=r_{f}\exp
\left[ {\Delta\theta U_{0}^{2}({U_{0}\over\sigma}+\sin^{2}\theta)
\over \sin 2\theta} \right]
\,.
\label{eq2}
\end{equation}
where $r_{f}$  is a lower limit of the integration. It may be seen
that a large factor $U_{0}^{3}/\sigma \gg 1 $ is present in
the expression above for small angles
 $\theta  < \sqrt{U_{0}/\sigma}$
and an even larger multiplier  $U_{0}^{2}$ at larger angles.
Collimation is firstly expected at small angels $\Delta\theta
\sim \theta$. Therefore, the distance at which the jet is formed is

\begin{equation}
r_{coll}=r_{f}\exp{{U_{0}^{3}\over 2\sigma}}.
\label{eq3}
\end{equation}
For the parameters used in the calculations  $r_{coll}=3.5 \times 10^{19} r_{f}$.
It is interesting to estimate the distances at which we should expect
collimation
of the wind from the Crab pulsar. For this pulsar $\sigma=10^{6}$
$\gamma_{0}=10^{3}$ and thus we get $r_{coll}\approx 10^{226}$ cm, a value
much larger than the size of the cavity (0.1 pc) formed by the wind
before it terminates in the interstellar medium.\\

\setcounter{figure}{12}
\begin{figure}
\centerline{\psfig{file=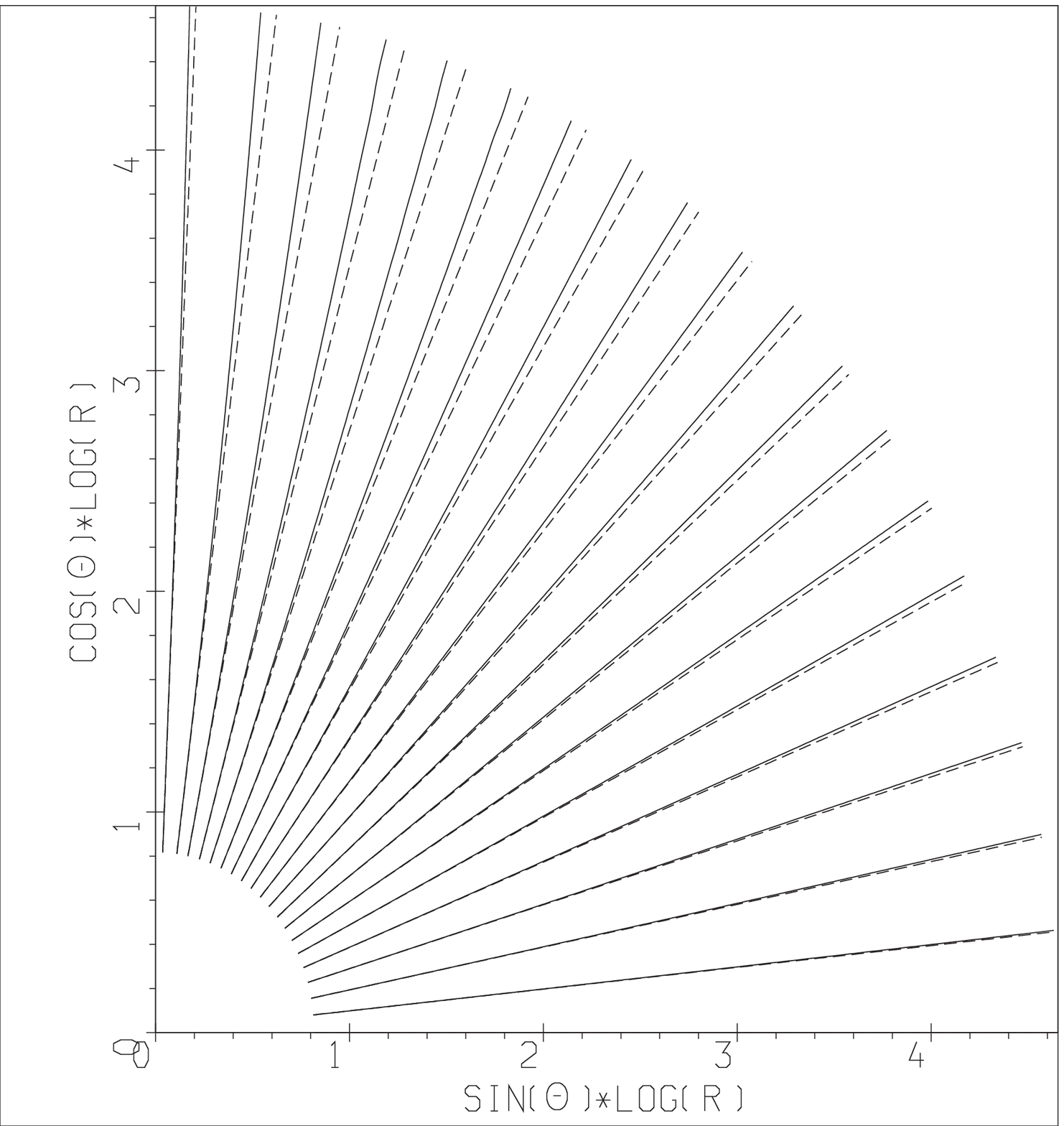,height=9.0truecm,angle=360}}
\caption{\it Shape of the poloidal magnetic field lines (solid) in the far
zone of a rotating magnetic rotator ejecting relativistic plasma.
Dashed lines show pure radial outflow.}
\label{relativ}
\end{figure}

This result shows that for parameters typical for radio pulsars
there is no collimation of the supersonic plasma at resonable distance.
Therefore, the plasma is not accelerated in this region as it was also 
concluded by Begelman \& Li (1994).

\section{Summary}

The axisymmetric 3-D MHD outflow of plasma from a magnetized and rotating
central object is numerically simulated for a wide range of angular velocities
of the central star. The simulation of the time-dependent evolution of
the flow from some initial state was used to obtain finally a stationary
solution. The model of a non rotating star with a monopole-like magnetic
field was taken as the initial state of the magnetosphere. It was found that
the obtained stationary final state depends critically on a single parameter
only. This parameter $\alpha$ expresses the ratio of the corotating speed at
the Alfv\'en distance to the initial flow speed along the magnetic
monopole-like fieldlines.
The acceleration of the flow was most effective at the equatorial plane and
the terminal flow speed depended linearly on $\alpha$.
Significant flow collimation was found in fast magnetic rotators corresponding
to large values of $\alpha > 1$, while very weak collimation occurs
in slow magnetic rotators with small values of $\alpha < 1$.
Part of the flow around the rotation and magnetic axis is cylindrically
collimated while the remaining equatorial part obtains radial asymptotics.
The transverse radius of the jet is found to be inversely proportional to
$\alpha$ while its density grows linearly with $\alpha$.
For $\alpha > 5$ the magnitude of the speed of the flow in the jet remained
below the fast MHD wave speed everywhere.  We predict that a regime of
nonstationary jet ejections { may be possible} at such high values 
of $\alpha$.

{ The above results have been obtained under several simplifying assumptions, 
such as, the neglect of gravity and thermal pressure as well as by taking 
for the initial magnetosphere a split-monopole configuration. Despite 
those assumptions however, we recover the main results of rather general 
theoretical studies on the formation of a collimated outflow from a 
magnetized and rotating astrophysical object. Nevertheless, for a meaningful 
comparison with the observations one needs to relax these assumptions, a task 
taken up in the next paper.}\\

{\bf Acknowledgements}. S.V. Bogovalov is grateful to the University of Crete
for financial support of his visit during collaborative work on this problem
and to K. Tsinganos for warm hospitality. BSV is also grateful to the
director of the Astrophysics Institute of MEPhI Yu. Kotov
for support of this work in Moscow.
Work of BSV was partially supported also by RFBR grant N 96-02-17113.
This research has been supported in part by a PENED grant from the General 
Secretariat for Research and Technology of Greece.

\end{document}